\documentclass[useAMS,usenatbib]{mn2e}
\usepackage{graphicx}

\def\url{Url:}

\title[T-P Profile for HD~189733b]{Temperature-Pressure Profile of the hot Jupiter HD~189733b from HST Sodium Observations: Detection of Upper Atmospheric Heating}
\author[C. M. Huitson et al.]{C. M. Huitson$^{1}$\thanks{E-mail: chuitson@astro.ex.ac.uk (CMH)}, D. K. Sing$^{1}$, A. Vidal-Madjar$^{2}$, G. E. Ballester$^{3}$,
\newauthor  A. Lecavelier des Etangs$^{2}$, J.-M. D\'esert$^{4}$, F. Pont$^{1}$ \\ 
$^{1}$Astrophysics Group, School of Physics, University of Exeter, Stocker Road, Exeter, EX4 4QL, UK \\
$^{2}$CNRS, Institut d'Astrophysique de Paris, UMR 7095, 98bis boulevard Arago, 75014 Paris, France \\
$^{3}$Lunar and Planetary Laboratory, University of Arizona, Sonett Space Sciences Building, Tucson, Arizona 85721-0063, USA \\
$^{4}$Harvard-Smithsonian Center for Astrophysics, 60 Garden Street, Cambridge, MA, 02138 \\}
\begin{document}

\date{Accepted 20 Feb 2012}

\bibliographystyle{mn2e}

\pagerange{\pageref{firstpage}--\pageref{lastpage}} \pubyear{2011}

\maketitle

\label{firstpage}

\begin{abstract}

We present transmission spectra of the hot Jupiter HD~189733b taken with the Space Telescope Imaging Spectrograph aboard \textit {HST}. The spectra cover the wavelength range 5808 - 6380~\AA $ $ with a resolving power of $R=5000$. We detect absorption from the Na~I doublet within the exoplanet's atmosphere at the $9\sigma$ confidence level within a 5~\AA $ $ band (absorption depth $0.09 \pm 0.01$ \%) and use the data to measure the doublet's spectral absorption profile. We detect only the narrow cores of the doublet. The narrowness of the feature could be due to an obscuring high-altitude haze of an unknown composition or a significantly sub-solar Na~I abundance hiding the line wings beneath a H$_2$ Rayleigh signature. These observations are consistent with previous broad-band spectroscopy from ACS and STIS, where a featureless spectrum was seen. We also investigate the effects of starspots on the Na I line profile, finding that their impact is minimal and within errors in the sodium feature.

We compare the spectral absorption profile over 5.5 scale heights with model spectral absorption profiles and constrain the temperature at different atmospheric regions, allowing us to construct a vertical temperature profile. We identify two temperature regimes; a $1280 \pm 240$~K region derived from the Na~I doublet line wings corresponding to altitudes below $\sim 500$~km, and a $2800 \pm 400$~K region derived from the Na~I doublet line cores corresponding to altitudes from $\sim 500-4000$~km. The zero altitude is defined by the white-light radius of $R_{\mathrm{P}}/R_{\mathrm{\star}}=0.15628 \pm 0.00009$. The temperature rises with altitude, which is likely evidence of a thermosphere. 

The absolute pressure scale depends on the species responsible for the Rayleigh signature and its abundance. We discuss a plausible scenario for this species, a high-altitude silicate haze, and the atmospheric temperature-pressure profile that results. In this case, the high altitude temperature rise for HD~189733b occurs at pressures of $10^{-5}$-$10^{-8}$ bar.

\end{abstract}


\section{Introduction}
\label{intro}

Transiting planets allow us to study exoplanet atmospheres through transmission spectroscopy, as the light from the host star passes through the planet's atmosphere. Hot Jupiters offer the best opportunity for these studies due to their large size and often large atmospheric scale heights.

Theory predicts the presence of strong Na~I~D lines in hot Jupiter atmospheres \citep{seagersasselov00,brown01}, and observations have so far agreed, with atomic sodium detected in hot Jupiters \citep{charbonneau02,redfield08,snellen08,sing08,wood11,jensen11}. However, the spectral shape of the sodium doublet absorption has been measured only for very few planets. The doublet samples a large range of altitudes in the atmosphere and by measuring line the absorption profile we can derive atmospheric temperatures, pressures and atomic sodium abundances. Those doublet profiles that have been measured show very different characteristics. The sodium absorption line profile in HD~209458b shows a narrow absorption core and broader plateau-like shape rather than smoothly decaying wings \citep{sing08,sing08b}. As explained by \citet{vidalmadjar11}, the absorption depth should be greater towards smaller bandwidths due to an increase of the absorption cross-section. A plateau region could indicate an abundance drop where the increase in the cross-section is compensated for by the decrease in atomic sodium abundance in the altitude regions sensed by narrower bands. The observation of pressure-broadened line wings shows that the Na I line is observed deep in the atmosphere and suggests a relatively cloud or haze-free atmosphere. 

Further analysis of transit observations of HD~209458b by \citet{vidalmadjar11} and \citet{vidalmadjar11b}, combining archive \textit{Hubble Space Telescope (HST)} and ground-based measurements \citep{snellen08}, allowed a temperature-pressure ($T$-$P$) profile to be constructed from $10^{-3}$ to $10^{-7}$~bar. The shape of the line absorption is characteristic of the atmospheric temperature profile, as higher temperatures increase the scale height thus causing a steeper line slope. \citet{vidalmadjar11b} found that the temperature rose with altitude over the range $10^{-3}$-$10^{-7}$ bar to $3600 \pm1400$~K at the highest altitudes, a sign of the planet's thermosphere. Models suggest that very hot thermospheres should be common in close-in extrasolar planets, due to the large amounts of UV radiation received from the nearby star. \citet{lecavelier04} and \citet{yelle04}, modelling the upper atmosphere and atmospheric escape of HD~209458b, showed that temperatures in the thermosphere could reach up to 10,000~K. Later models from \citet{tian05} and \citet{garciamunoz07} showed temperatures of a similar order of magnitude.

In contrast to the measured profile for HD~209458b, WASP-17b appears to show only a narrow absorption core, with no observable line wings \citep{wood11}. This indicates that there could be absorbing species very high in the atmosphere, obscuring the line wings.

For HD~189733b, broad-band spectroscopy has shown a spectrum dominated by Rayleigh scattering from 3000 to 10500~\AA $ $ \citep{pont08,sing09,sing11,gibson12}, with longer wavelengths probing deeper in the atmosphere. This featureless broad-band spectrum likely indicates that the Na~I line wings produced by pressure broadening deep in the atmosphere are obscured by continuum absorption higher in the atmosphere. Na~I has been detected in HD~189733b from the ground, with a relative absorption depth of $(67.2 \pm 20.7) \times 10^{-5}$ \citep{redfield08} compared to the continuum. This detection was later revised to $(52.6 \pm 16.9) \times 10^{-5}$ by \citet{jensen11}. 

An upper atmospheric temperature-pressure profile has only been derived for one exoplanet, HD~209458b. The aim here is to conduct a similar study for HD~189733b. These two planets are in many respects at opposite ends of the spectrum of hot Jupiters. HD~209458b has a bloated radius, displays a stratospheric temperature inversion and orbits a star that is much less active than HD~189733. In contrast, HD~189733b does not have a bloated radius, shows no stratospheric temperature inversion and orbits one of the most active extrasolar planet host stars. HD~189733 is a somewhat cooler star than HD~209458 (4980 vs 6075~K). Additionally, the transit absorption spectrum of HD~209458b is dominated by atomic absorption from alkali lines, whereas that of HD~189733b appears dominated by a Rayleigh scattering haze. \citet{heng11} demonstrate that the presence of clouds and hazes can have very different effects on the temperature-pressure profile depending on their thicknesses and opacities. IR absorption due to a haze layer can heat the lower atmosphere and cool the upper atmosphere, whereas optical scattering will have the opposite effect, warming the upper atmosphere. The many differences between the two planets makes HD~189733b an ideal planet to study in detail for comparison with HD~209458b.

Since many factors can affect the energy budget of a planetary atmosphere, such as advection efficiencies and opacities in the atmosphere, it is difficult to say in advance how incoming stellar radiation on HD~189733b affects the atmospheric temperature profile. Differing levels of radiation from the host stars may have a significant effect on the upper atmospheric chemistry as well. \citet{knutson10} suggest that planets that orbit the most active stars have non-inverted stratospheres and planets that orbit quiet stars do have inverted stratospheres. The authors suggest that the excess EUV radiation from a more active star can break down the compounds responsible for stratospheric temperature inversions, thus altering the $T$-$P$ profile significantly even at altitudes lower than the regime observed here. 

In this paper, we present \textit{HST} Space Telescope Imaging Spectrograph (STIS) medium-resolution transmission spectra of HD~189733b with the G750M grating. The increased resolution and high signal-to-noise of these spectra compared to the ground-based detection allows us to measure the spectral absorption depth profile of the sodium doublet and construct a vertical temperature profile of the atmosphere. Making an assumption of the pressure at a reference altitude, this translates into a temperature-pressure profile which can be compared to theoretical predictions.  

\section{Observations and Data Reduction}
\subsection{Light Curve Fitting}
\label{sec_light_curves}

We observed 7 primary transits of HD~189733b using the \textit{HST} STIS G750M grating (\textit{HST} GO 11572, PI: David Sing). Three of these transits obtained data of high quality. Of the others, two suffered from a positioning error of the spectrum when using sub-arrays on the CCD, as also noted in \citet{brown01b}. Two further visits obtained no data as those visits suffered from guide star problems. Table \ref{table_visits} gives details of the 7 visits.

\begin{table}
\centering
  \begin{tabular}{c | c | c | c}
\hline
Visit & JD & Notes \\
\hline
1 & 2455135.31 & Positioning error. Spectrum blueward of   \\
& & 6000~\AA $ $ used in part of the analysis \\
2 & 2455141.42 & No data acquired \\
3 & 2455148.62 & Whole spectrum useable \\
4 & 2455184.11 & No data acquired\\
5 & 2455423.72 & Positioning error. Spectra were \\
& & unuseable \\
6 & 2455470.31 & Whole spectrum useable \\
7 &  2455530.22 & Whole spectrum useable \\
\hline
\end{tabular}
\caption{Table of \textit{HST} STIS visits, showing data quality.}
\label{table_visits}
\end{table} 

Each dataset consists of a series of 144 spectra, each covering a wavelength range of 5808-6380~\AA. The G750M grating has a resolving power of $R=5000$, which gives a resolution of $\sim$ 1.2~\AA $ $ or $\sim 2$ pixels at 5890~\AA. The observations were made with a slit of 2\verb+"+  width to minimise light loss. The data were then bias-subtracted, dark-subtracted and flat-fielded using the CALSTIS pipeline. The spectra were extracted afterwards using IRAF with an aperture 13 pixels wide. IRAF was also used to remove cosmic ray contamination and to measure the spectral trace (the position of the spectrum on the CCD). These spectra were then shifted to a common wavelength scale by cross-correlating with the mean of the shifted spectra, and were then corrected to a rest frame by comparison with a model spectrum.

\begin{figure}
\centering
\includegraphics[width=8cm]{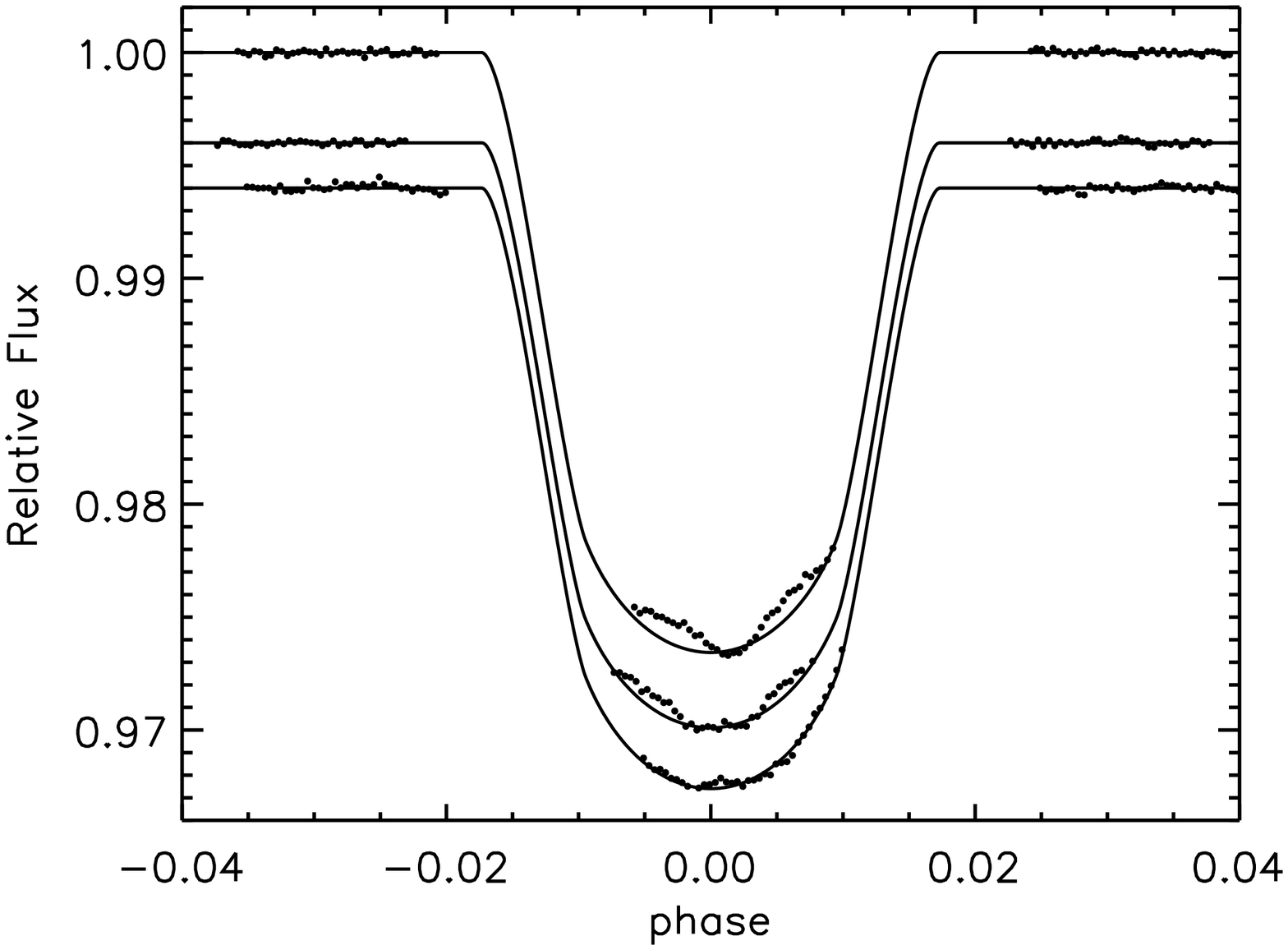}
\includegraphics[width=8cm]{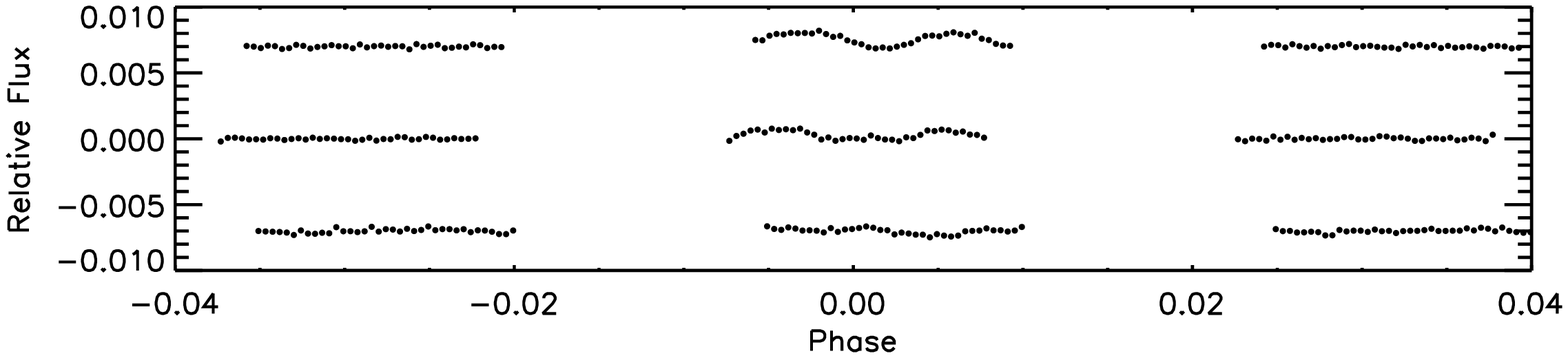}
\caption{White light curves for (top to bottom) visit 3, visit 6 and visit 7 with the systematic effects removed. An arbitrary flux offset has been applied for clarity. For each visit, the best fit analytical transit model of \citet{mandelagol02} is over-plotted. Plotted underneath are the residuals for (top to bottom) visit 3, visit 6 and visit 7 after the systematics have been removed. An arbitrary flux offset has been applied for clarity. The regions where the planet crosses stellar spots can be clearly seen.}
\label{fig_light_curves}
\end{figure}

Transit light curves were produced by summing the photon flux over the full spectrum in each exposure (see Figure~\ref{fig_light_curves}), and were modelled using the analytical transit models of \citet{mandelagol02}. The light curves exhibit all the instrumental effects noted by \citet{brown01b}. The systematic effects were corrected for by discarding the first orbit of each visit and the first exposure of the retained orbits, and then removing a fourth-order polynomial dependence of the fluxes with \textit{HST} orbital phase based on a fit to the out-of-transit exposures. The Levenberg-Markwardt least-squares technique was used for the fit using the IDL MPFIT package \citep{markwardt09} using the unbinned data. The in-transit sections obviously crossing stellar spots were removed in these fits.

We also investigated whether fitting for any remaining $x$ and $y$ offsets with of the spectra on the detector (such as due to target motion within the aperture) with linear functions were justified by improvement in the fit as done by \citet{sing11}. These variations improved the reduced $\chi^{2}$ and significantly improved the Bayesian information criterion (BIC), which penalises models with larger numbers of free parameters, and prevents over-fitting the data. Table \ref{stats} gives details of these improvements.

\begin{table}
\centering
\begin{tabular}{c|c|c|c|c|c|c}
\hline
& \multicolumn{2}{|c|}{Visit 3} & \multicolumn{2}{|c|}{Visit 6} & \multicolumn{2}{|c|}{Visit 7} \\
\hline
Fit Position & no & yes & no & yes & no & yes \\
Parameters? & & & & & \\
DOF & 69 & 66 & 79 & 76 & 86 & 83 \\
NFree & 8 & 11 & 8 & 11 & 8 & 11 \\
$\chi^2_{\nu}$ & 1.948 & 1.751 & 5.705  & 1.958 & 3.632 & 1.936 \\
BIC & 349 & 167 & 447 & 183 & 265 & 211 \\
\hline
\end{tabular}
\caption{Table of fitting statistics depending on whether the remaining $x$ and $y$ offsets were included in the fit. `NFree' is the number of free parameters. It can be seen that there are significant improvements for all the visits when these parameters are fitted.}
\label{stats}
\end{table} 

Fitting the systematic trends in this way produced S/N values for the white light curves between 10,000-11,000 (precision levels of 90-100 ppm), which are 78-86 per cent of the Poisson-limited value. These per exposure precision levels are an improvement on the results when not fitting for these additional position-dependent parameters (with precision levels of 130-160 ppm). 

We checked for red noise by observing whether residuals binned in time followed a $N^{-1/2}$ relationship, where $N$ is the number of points in the time bin. We found no significant evidence of red noise in visit 3 or visit 6 for the white light curve. For visit 7, the magnitude was $3 \times 10^{-5}$ for the white light curve, which corresponds to a S/N of $\sim 33,000$. It translates into an error in depth measurement in the planet's atmosphere of $\sim 50$~km, which is less than 1/4 of a scale height in the lowest regions (smallest scale height) probed by these observations. Since our uncertainties are $\sim 100-300$~km in altitude, these low levels of systematic trends do not dominate the uncertainties in our measured spectral absorption depth profile. Smaller bandwidths are generally closer to Poisson-limited, and it was also found that the red noise in a 20~\AA $ $ band was negligible ($< 1 \times 10^{-5}$). 

We also used linear regression to check for remaining correlation between fitted parameters, and all correlations were of very small order (in the range 0.0008-0.0001). Figure~\ref{fig_light_curves} also shows the residuals after removing the polynomial characteristic of the systematic effects. Visits 3 and 6 still exhibit structure in the transit due to occulted starspots, where the flux is higher than the transit model predicts (see Section~\ref{spots}). It was observed that the transit depth and hence the sodium absorption depths for visit 6 was shallower than the other two visits. This could be due to an occulted starspot in the centre of the transit in addition to those at the edges, so that the planet would cover part or all of at least one starspot during the whole transit. 

\subsection{Stellar Limb Darkening}

Correcting for the effects of stellar limb darkening on the transit light curves and the spectral absorption profiles is important at visible wavelengths. We compared two limb darkening corrections based on two stelar models; the Kurucz (1993) 1D ATLAS stellar atmospheric model \footnote[1]{See http://kurucz.harvard.edu/stars/hd189733.} and a fully 3D time-dependent hydrodynamic stellar atmospheric model (\citealt{sing11, hayek12}). The models were compared using visit 7, where there seem to be no occulted spots at the limbs of the transit. 

In the 1D case, \citet{sing09} and \citet{sing10} found that a four-parameter law broke down for small $\mu$, where $\mu = \cos(\theta)$, and $\theta$ is the angle in radial direction from the disc centre. Since the first coefficient has a greater effect at small angles, it was not used, and we instead fitted a 3 parameter law:

\begin{equation}
\frac{I(\mu)}{I(1)}=1-c_{2}(1-\mu)-c_{3}(1-\mu^{3/2})-c_{4}(1-\mu^2) .
\label{fourth_order_ld}
\end{equation}
The stellar parameters used were $T_{\mathrm{eff}}=5000$~K, $\log (g)=4.5$, and [Fe/H]$=0.0$, along with the transmission function of the G750M grating. The resulting coefficients for white light were $c_{2}=0.8918$, $c_{3}=0.0515$ and $c_{4}=-0.1930$. This fit gave a reduced $\chi^2$ of 2.52 and BIC of 268.

In the 3D case, the stellar limb darkening was fitted with a 4-parameter law, as it appears to perform well at small $\mu$:

\begin{equation}
\frac{I(\mu)}{I(1)}=1-c_{1}(1-\mu^{1/2})-c_{2}(1-\mu)-c_{3}(1-\mu^{3/2})-c_{4}(1-\mu^2) .
\label{eqn_3d_ld}
\end{equation}
The parameters for white light were $c_{1}=0.7043$, $c_{2}=-0.4493$, $c_{3}=1.0538$, and $c_{4}=-0.4569$. It was found that the fits and the magnitudes of red noise improved over the 1D case. The reduced $\chi^2$ when using the 3D model was 1.94, with a BIC of 211. Despite this, however, the measured planet-to-star radius changed by less than $1 \sigma$. \citet{sing11} give an in-depth discussion of the comparison between the 1D and 3D cases for broadband G430L transit observations (2900-5700~\AA). They also find that the fits to their data are significantly better using the 3D model than the 1D model. 

\section{Analysis}
\label{analysis}

\subsection{Spectral Absorption Depth Profile of the Na~I Doublet}
\label{na_spectral_profile}

We measured the spectral absorption depth profile versus wavelength around the sodium lines by comparing the absorption depths in the sodium wavelengths to absorption depths in a reference band during transit. For the reference band, we used the `wide' band defined by \citet{charbonneau02}, which is a combination of a blue (5818-5843~\AA) and a red (5943-5968~\AA) band. The three complete visits were used for this analysis, as was visit 1, since we only required wavelengths up to 5968~\AA. We did not use the fitted radius from visit 1, only the differential flux measurement described here, so this visit is not discussed in Section~\ref{sec_light_curves}.  We tested using the `medium' reference band as defined by Charbonneau et al. (2002) as well as a reference band closer to the doublet, and found that the choice of reference band shifts the profile in altitude by less than $\frac{1}{2} \sigma$.

This differential method has the advantage that systematic errors largely cancel out, as they are largely wavelength independent, giving a robust way to measure Na~I absorption. It also averages the blue and red background components, giving an average that largely compensates for the effects of limb darkening in the sodium bands. This method is also particularly useful here due to the presence of occulted starspots at the limbs of two of the visits, as the flux variations due to transiting in front of occulted spots are similar in the Na band and the `wide' bands, and thus cancel out when comparing the bands.

However, this method does not account for differential limb darkening in the sodium line core wavelengths. The limb darkening is weaker at the line core compared to the continuum, because the core of the stellar Na I line is produced above the photosphere, further away from the centre of the star than where the continuum is produced. At these altitudes, the temperature gradient becomes smaller (see Hayek et al., submitted). The differential limb darkening leads to a slight underestimation of the absorption depth in the line core, since the limb darkening effects are not fully cancelled out by the differential method. This effect was corrected by comparing limb darkening models for each of the evaluated sodium wavelengths with that of the `wide' band. These models were given a fixed planetary radius, to evaluate the difference in model limb darkening, and then the differences were added to the sodium absorption depths. The effects were found to be negligible for wavelengths greater than 3 \AA $ $ away from the line cores.

\begin{figure}
\centering
\includegraphics[width=8cm]{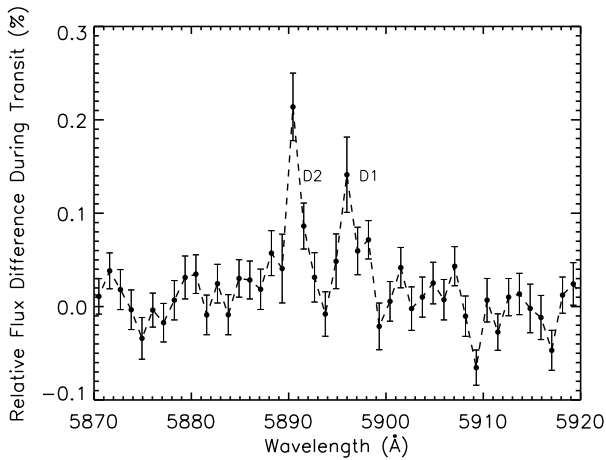}
\caption{Spectral absorption depth profile at each wavelength in the sodium region with 1$\sigma$ error bars, binned to the STIS resolution of 2 pixels. The Na~I~D1 and D2 doublet lines are resolved, as shown by the largest two peaks. The absorption depths for the doublet are greater than the equivalent profile for HD 209458b \citep{sing08}.}
\label{fig_sodium_spectrum}
\end{figure}

Figure~\ref{fig_sodium_spectrum} shows the spectral absorption depth profile around the Na~I feature binned to the instrument resolution (2 pixels) and Table \ref{table_absorption} lists the differential absorption depths for each wavelength bin. We find in Section~\ref{spots} that the effects of starspots on this profile are minimal and within our observational uncertainties. Comparing Figure~\ref{fig_sodium_spectrum} to the equivalent spectrum for HD~209458b from \citet{sing08} shows that the measured differential absorption for HD~189733b is deeper in the line cores. There is one data point in the centre of the two doublet lines which appears to have an unexpectedly low absorption in the unbinned data, and can still be seen in the binned data. It is not associated with any excess red noise, and so we could not scale the uncertainty any higher. This anomaly was also observed in the HD~209458b spectrum from \citet{sing08b} using the same instrument, but there are no obvious defects on the CCD that could account for this effect. 

\begin{table}
\centering
  \begin{tabular}{c | c | c | c | c | c}
\hline
$\lambda$ & AD & Error & $\lambda$ & AD & Error \\
 (\AA) & ($\times 10 ^{-5}$) & ($\times 10 ^{-5}$) & (\AA) & ($\times 10 ^{-5}$) & ($\times 10 ^{-5}$) \\
\hline
5870.50 &  10.95 &  19.04 & 5871.61 &  38.27 &  19.24 \\
5872.71 &  18.20 &  21.41 & 5873.82 &  -3.38 &  21.21 \\
5874.93 & -34.05 &  22.34 & 5876.04 &  -3.78 &  18.22 \\
5877.15 & -17.25 &  20.63 & 5878.25 &   6.80 &  21.09 \\
5879.36 &  31.16 &  23.01 & 5880.47 &  34.76 &  20.70 \\
5881.58 &  -8.91 &  21.23 & 5882.69 &  24.54 &  20.73 \\
5883.79 &  -8.71 &  21.49 & 5884.90 &  30.13 &  20.06 \\
5886.01 &  28.46 &  20.34 & 5887.12 &  18.51 &  21.61 \\
5888.23 &  57.24 &  24.06 & 5889.33 &  40.69 &  36.82 \\
5890.44 & 213.89 &  36.17 & 5891.55 &  86.23 &  24.62 \\
5892.66 &  31.31 &  26.42 & 5893.77 &  -7.95 &  23.81 \\
5894.87 &  48.48 &  29.33 & 5895.98 & 141.22 &  40.31 \\
5897.09 &  59.56 &  25.44 & 5898.20 &  71.53 &  20.53 \\
5899.31 & -21.23 &  25.01 & 5900.41 &   5.60 &  21.23 \\
5901.52 &  41.71 &  21.60 & 5902.63 &  -2.19 &  23.29 \\
5903.74 &  10.17 &  21.49 & 5904.85 &  25.18 &  22.27 \\
5905.95 &   7.36 &  21.65 & 5907.06 &  43.25 &  20.97 \\
5908.17 & -10.27 &  21.61 & 5909.28 & -65.38 &  18.88 \\
5910.39 &   6.79 &  23.31 & 5911.49 & -27.29 &  19.54 \\
5912.60 &  10.12 &  19.95 & 5913.71 &  13.50 &  22.14 \\
5914.82 &  -2.03 &  26.01 & 5915.93 & -11.71 &  23.66 \\
5917.03 & -46.86 &  21.42 & 5918.14 &  12.27 &  19.40 \\
5919.25 &  24.25 &  22.80 & & & \\

   \hline
\end{tabular}
\caption{Weighted average of spectral absorption depth (AD) in each wavelength ($\lambda$) around the Na~I doublet compared to absorption in adjacent bands. The data are binned to the STIS instrument resolution of 2 pixels.}
\label{table_absorption}
\end{table}

\subsection{Integrated Absorption Depth Profile of the Na~I Doublet}
\label{na_absorption_profile}

In order to allow direct comparison with the previously published Na~I profiles of HD~209458b and WASP-17b \citep{sing08,wood11}, we also compute the differential integrated absorption depth profile as a function of bandwidth. Using the method described by \citet{charbonneau02}, the absorption depth during transit in a band centred on the Na~I doublet was compared the same blue and red reference bands as used for the spectral absorption depth profile. The bands centred on each sodium doublet line (5896 and 5890 \AA) were then increased in size from 3 to 80~\AA, with a single band being used where the bands were large enough to encompass both lines together (bandwidth $\ge 12$ \AA). We heareafter refer to this profile as the integrated absorption depth profile.

\begin{figure}
\centering
\includegraphics[width=8cm]{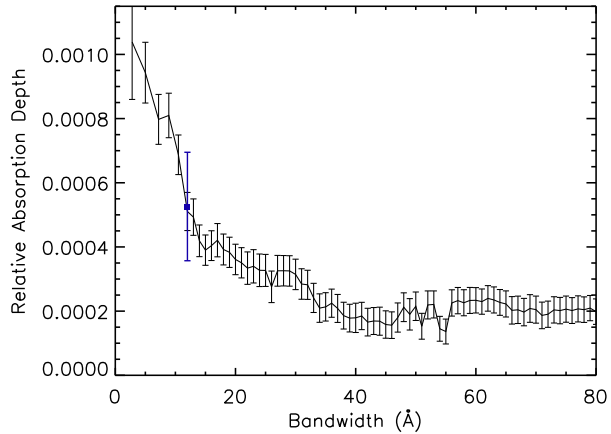}
\caption{The integrated absorption depth profile comparing the absorption depth in the Na~I doublet to the absorption depth in adjacent bands as a function of bandwidth around the sodium doublet. The detection from \citet{jensen11} is shown in blue with a filled square. Our more precise results agree with their measurement at the 1~$\sigma$ level.}
\label{fig_abs_vs_band}
\end{figure}

Figure~\ref{fig_abs_vs_band} shows the integrated absorption depth profile as a weighted average for the 4 visits. The line appears wider than in reality due to a dilution effect of increasing the bandwidth, weakening the sodium absorption signature confined to smaller bandwidths. The profile gives a depth of $(51.1 \pm 5.9) \times 10^{-5}$ for a bandwidth of 12~\AA, which was the bandwidth used by \citet{redfield08} and \citet{jensen11} for their ground-based measurements. Our result is within $1~\sigma$ of these previous results, confirming the absorption depth found with an increased S/N of almost a factor of 3.

\subsection{The Effect of Occulted Starspots}
\label{spots}

The local flux increases seen as bumps in the light curves in Figure~\ref{fig_light_curves} are most likely due to occulted starspots. HD~189733 is a very active star, showing periodic flux variations over time, and starspots are clearly seen in other transit light curves as well \citep{sing11,pont08,henrywinn08}. Since starspots are darker than the surrounding stellar surface, the planet blocks less of the stellar light when crossing a spot than when crossing a non-spotted region of the surface, resulting in a local flux increase in the light curve compared to the expected model. Occulted spots could therefore have an effect on the measured sodium absorption depth because they decrease the inferred planetary radius. 

\begin{figure}
\centering
\includegraphics[width=8cm]{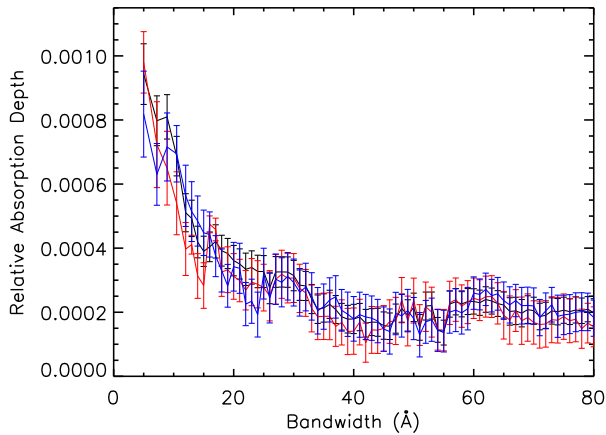}
\caption{Weighted average of integrated absorption depth in a band centred on the Na~I doublet compared to absorption in adjacent bands. \textit{Black: }using all exposures, \textit{blue: }using only exposures clearly not contaminated by occulted starspots, \textit{red: }using only exposures that appear to contain occulted starspots. This shows that the starspots do not have a significant effect on the Na~I absorption profile measurement at our level of precision. A colour version is available in the online journal.}
\label{fig_abs_vs_band_spot}
\end{figure}

The differential method of measuring the absorption profile should mostly cancel out the effects of stellar spots. However, we do not a priori know how much the level of unocculted stellar spots on the surface affects the stellar sodium signature, and hence the effect on the inferred planetary absolute absorption depth. This effect could be greater in the line core than in the broad regions of the line and could be significant if the sodium signature is strongly dependent on the spot temperature. The sodium data here are used to empirically determine the effect of occulted spots on the differential Na~I spectral absorption depth profile. We perform this analysis on the integrated absorption depth profile because the effects will be more severe in this profile than in the spectral absorption depth profile since they will be most pronounced in the line core. Figure~\ref{fig_abs_vs_band_spot} shows the mean integrated absorption depth versus bandwidth for the 4 visits compared with the mean integrated absorption depths evaluated using only the exposures that appear to occult stellar spots, and also only those that do not.

As expected, the derived integrated absorption depths are smaller for when evaluated from exposures occulting spots, with a value of $(39.7 \pm 8.2) \times 10^{-5}$ integrated over a 12~\AA $ $ band, compared to $(56.3 \pm 9.6) \times 10^{-5}$ derived from the exposures not occulting spots. The effect at even smaller bandwidths, where we expect the difference to be more noticeable, is around $5-15 \times 10^{-5}$, still within our observational error bars. For our purposes, we can therefore assume that the spots have a similar spectrum to the non-spotted surface. In this case, unocculted starspots are expected to have an even less significant effect than occulted starspots, so we can assume that these effects are also smaller than our uncertainties in absorption depth. It has, however, been noticed that at both longer and shorter wavelengths in the broad-band data, the effects become more significant \citep{sing11}. 

\subsection{Broadband 5808-6300~\AA $ $ Spectrum}

The G750M data can also be used to look for other spectral features and to determine the shape of the overall spectrum from 5808 to 6380~\AA. Spectra were created by binning the spectral time series into 5 wavelength bins of $\sim 110$~\AA $ $ for visits 3, 6 and 7. Each band was fitted individually in the manner described for the white light curve. The fitted planet to star radius ratios were then used to construct a binned transmission spectrum. 

\begin{figure}
\centering
\includegraphics[width=8cm]{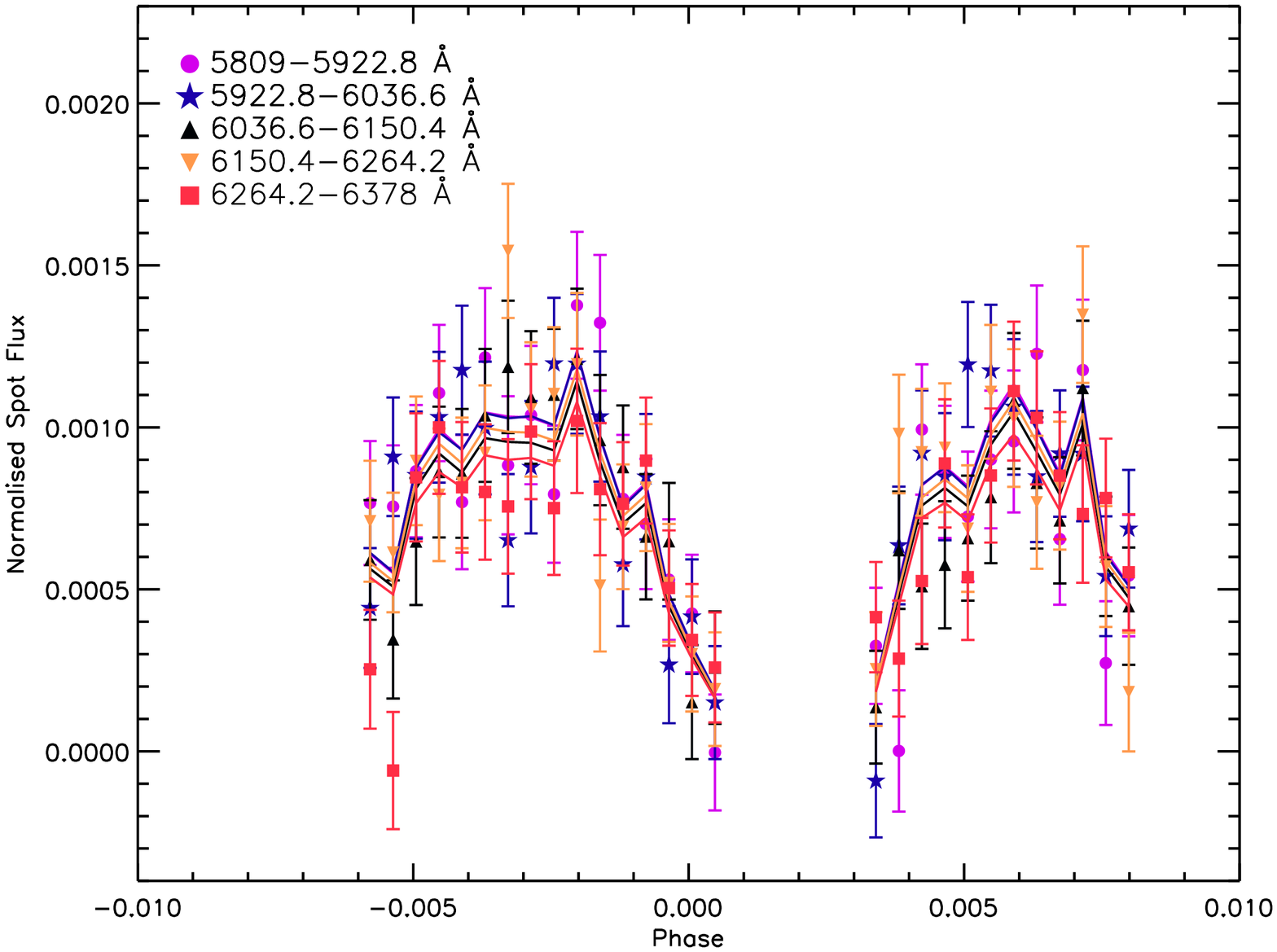}
\includegraphics[width=8cm]{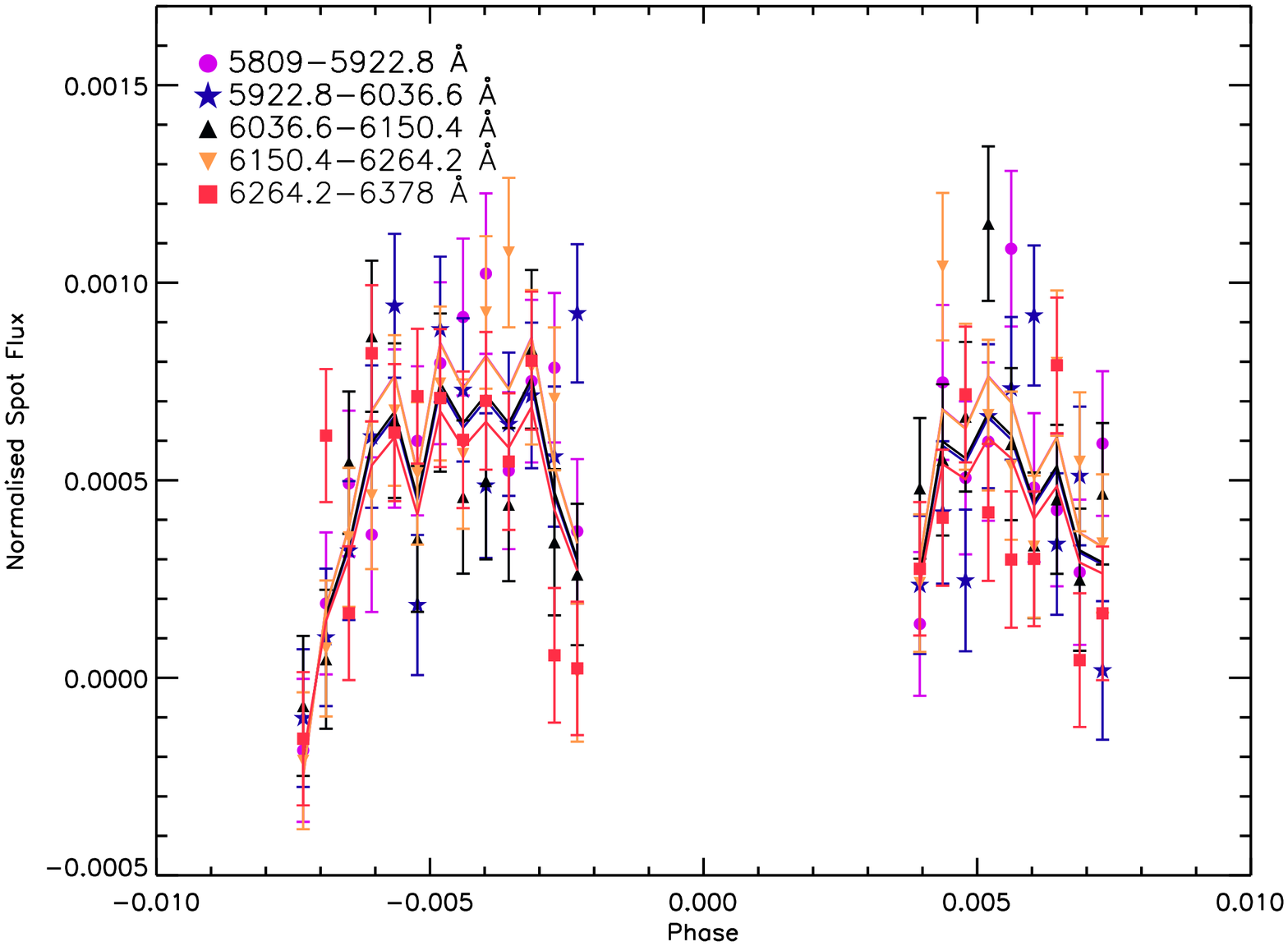}
\caption{Transit exposures sampling stellar spots overplotted with the best fit spot model for visits 3 (top) and 6 (bottom). A non-spotted model has been subtracted from both. A colour version is available in the online journal.}
\label{fig_spots}
\end{figure}

\begin{figure}
\centering
\includegraphics[width=8cm]{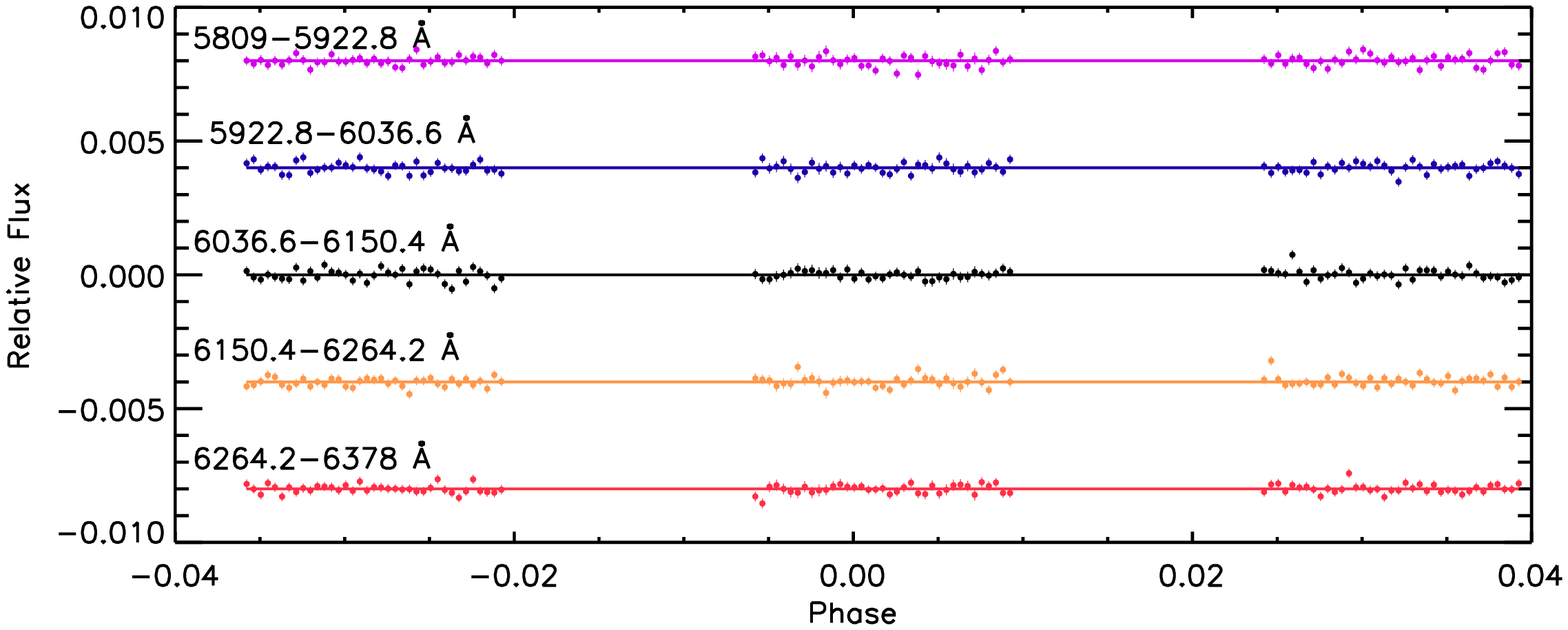}
\includegraphics[width=8cm]{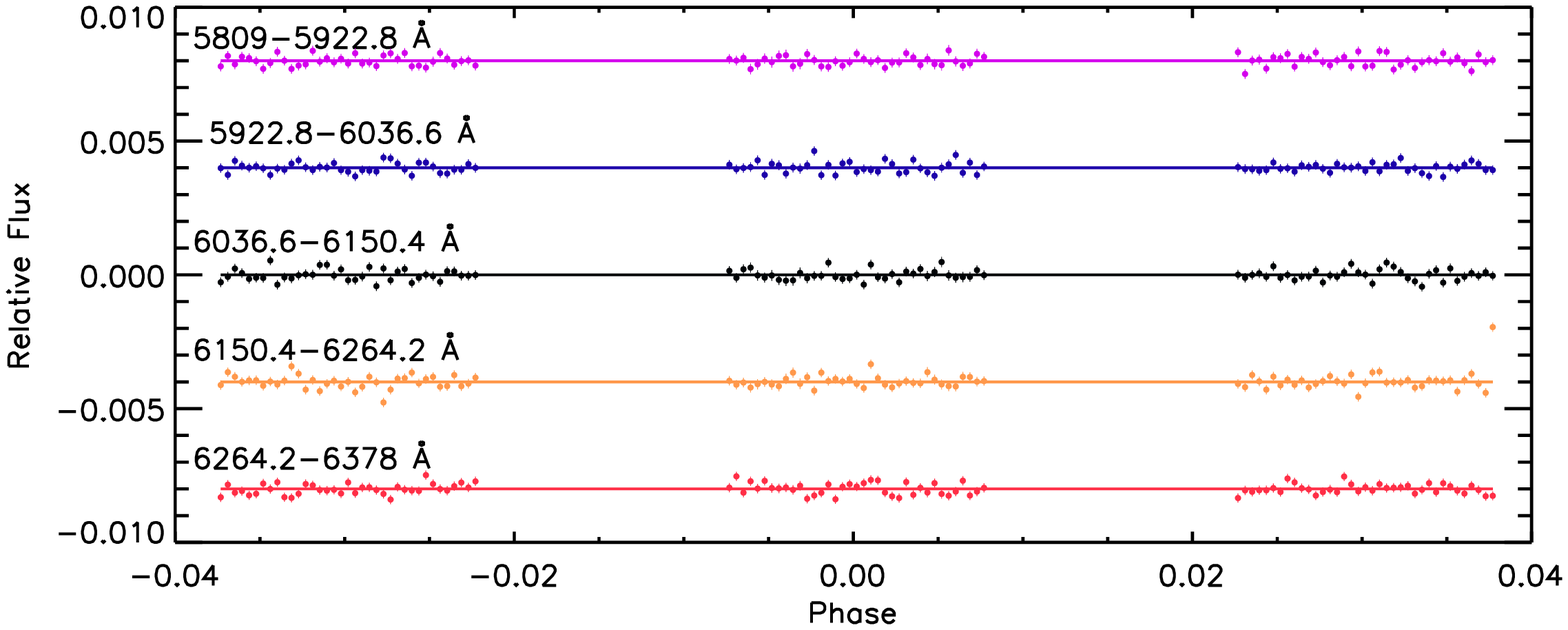}
\caption{Residuals of the data minus spotted models for visit 3 (top) and visit 6 (bottom) for the G750M band divided into 5 $\sim 110$ \AA $ $ bins. Arbitrary flux offsets have been applied for clarity. The colours and wavelength bands are the same as in Figure~\ref{fig_spots}, and a colour image can be found in the online journal. The flatness of the residuals show that the occulted starspots have been effectively removed.}
\label{fig_nonspot_residuals}
\end{figure}

Determining accurate radii required the light curves to be corrected for occulted stellar spots. The fluxes of each exposure contaminated by occulted spots in visits 3 and 6 were measured with the white light curve, producing the shape of the features. These were then used to model the spots in each of the wavelength bands, by fitting the features in each band with the shape measured from the white light curve and an amplitude parameter. This assumes that the shape of the spots is roughly the same in all the different bands and that the spectra of the spots are constant. Figure~\ref{fig_spots} shows the spotted exposures overplotted with the best fit spot amplitudes for each wavelength band. Figure~\ref{fig_nonspot_residuals} shows the data residuals once the spots have been removed. Comparison with Figure~\ref{fig_light_curves} shows that the spots have been effectively removed.

\begin{figure}
\centering
\includegraphics[width=8cm]{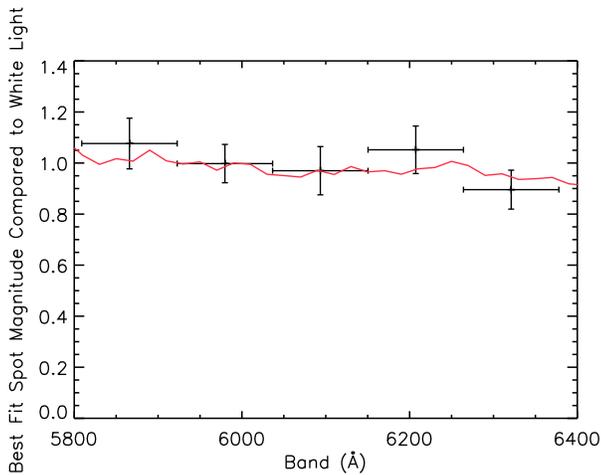}
\caption{Average best fit spot magnitudes from the models for visit 3 and visit 6, plotted against band. It can be seen that there is a wavelength-dependent slope, but this is within the errors of the measured parameters. Overplotted in red is a model spectral signature of occulted spot amplitude, assuming that the spots have a temperature of 4000~K. The measured slope is in agreement with this model, but cannot be used to constrain the temperature any further. }
\label{fig_spotfluxes}
\end{figure}

Starspots are cooler than the surrounding surface, and so are comparatively brighter at longer wavelengths. Occulted spot features are therefore more severe at blue wavelengths than at red (e.g. \citealt{pont07}). Figure~\ref{fig_spotfluxes} shows the best-fit spot amplitude parameters along with a predicted slope for spots of temperature 4000~K. \citet{sing11} found that this temperature best matched their data over a broad wavelength range. The amplitude was defined to be 1 for white light. Our slope is in agreement with this model, but cannot be used to constrain the spot temperature further than the constraints provided by previous work. It can be seen that the amplitude increases slightly towards blue wavelengths as expected, although this is only a $1~\sigma$ effect over our wavelength range. 

\begin{figure}
\centering
\includegraphics[width=8.5cm]{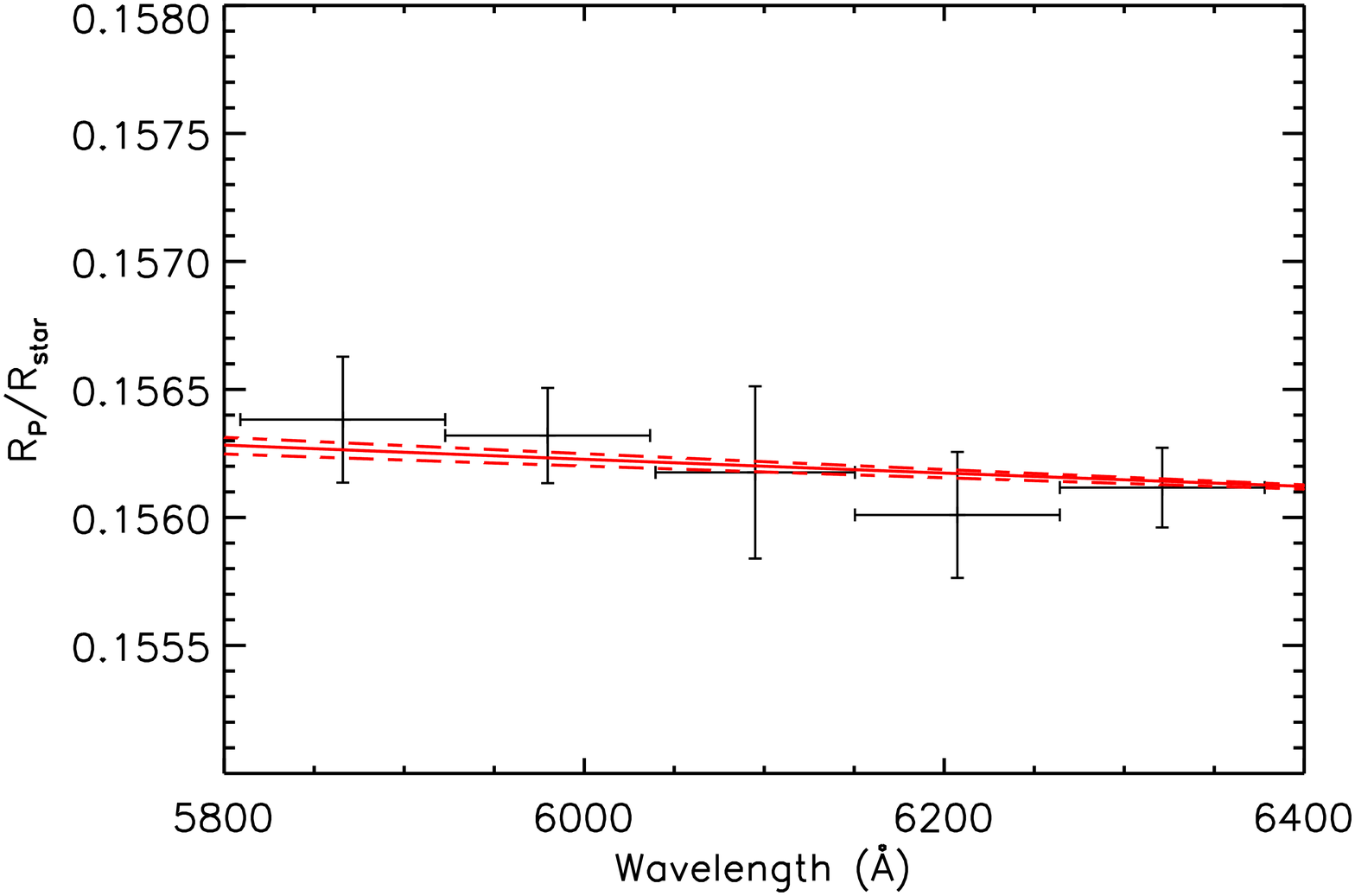}
\caption{The STIS G750M broad spectrum overplotted with a Rayleigh scattering prediction from the ACS data \citep{pont08,sing11} at the temperature of 1340~K determined by \citet{lecavelier08} (solid line) with a range of $\pm 150$~K (dashed lines).}
\label{broad_spectrum}
\end{figure}

The planetary radius and spot magnitude parameter were initially fitted together and then fitted one at a time iteratively until the solutions converged, allowing the radii in each wavelength band to be fitted while taking into account the shape of the spots in the light curve. The spectrum was then corrected for unocculted spots by reducing the measured radii by 1\% (see \citealt{sing11}) and averaged for the 3 visits. Figure~\ref{broad_spectrum} shows the resulting spectrum and Table~\ref{table_broad_spectrum} shows the fitted values of $R_{\mathrm{P}}/R_{\star}$. Figure~\ref{broad_spectrum} also shows a Rayleigh slope for an atmospheric temperature of $1340 \pm 150$~K determined from previous observations \citep{lecavelier08}. Our data are consistent with Rayleigh type scattering, but we cannot further constrain the uncertainty on the measured temperature with our data, although we do reach precisions of approximately one atmospheric scale height for each of our binned spectral points. Additionally, we see a flat spectrum with spectral bins narrower than those of either of the two previously observed broad-band spectra, which is consistent with the conclusion that the Na~I line wings are absent. It also indicates that there are no other broad spectral features in our wavelength range. 

\begin{table}
\centering
  \begin{tabular}{c | c | c}
\hline
Wavelength (\AA) & $R_{\mathrm{P}}/R_{\star}$ & Error \\
\hline
5809-5922 & 0.15638 & 0.00025\\
5923-6037  & 0.15632 & 0.00019 \\
6040-6150 & 0.15618 & 0.00034 \\
6150-6264 & 0.15601 & 0.00025 \\
6264-6378 &  0.15612 & 0.00016 \\
   \hline
\end{tabular}
\caption{Fitted planet-to-star radius ratios for $\sim 110$ \AA $ $ bands, used to construct the broad-band spectrum for the G750M band. The radii have been corrected for occulted and unocculted stellar spots.}
\label{table_broad_spectrum}
\end{table}

In contrast to the flat spectrum measured here, the spectrum of HD~209458b showed significant excess absorption at 6200~\AA $ $ \citep{sing08,desert08}. The non-detection of significant excess absorption at 6200~\AA $ $ in HD~189733b suggests that, either this feature is real or any systematics responsible for the feature vary with time. 

\subsection{Determining Upper Atmospheric Temperatures}

Visible transmission spectroscopy and the sodium D doublet in particular should probe the upper atmosphere, at pressures below 150~mbar to as low as $10^{-9}$ bar, where we could see very high temperatures \citep{lecavelier04,yelle04, garciamunoz07,moses11}. \citet{vidalmadjar11, vidalmadjar11b} measured a temperature of $3600 \pm 1400$~K for the upper atmosphere of HD~209458b, using the narrowest bands (highest altitudes) of the exoplanet's sodium absorption profile. Here, we use the same method to determine the temperature ranges covered by our measurements. We generate models of the sodium D doublet absorption lines at various temperatures and calculate spectral absorption depth profiles at the spectral wavelengths of the measurements. This method is independent of absolute absorption depth and depends only on the absorption profile slope, which means that we do not have to know a reference pressure scale or the abundances to measure local temperatures. We assume constant sodium abundance, which is valid for local measurements.

The atmospheric altitude, $z(\lambda)$, that can be ascribed to the observed absorption as a function of wavelength in a transmission spectrum, is given by the equation from \citet{lecavelier08}:

\begin{equation}
z(\lambda)=\frac{k T}{\mu g}\textup{ln}\left ( \frac{\xi \sigma(\lambda) P_{o}}{\tau_{\mathrm{eq}}}\left ( \frac{2\pi R_{\mathrm{P}}}{k T \mu g} \right )^{1/2} \right ) ,
\label{eqn_guillot2}
\end{equation}
where $k$ is Boltzmann's constant, $T$ is the temperature, $\mu$ is the mean molecular weight of the atmospheric composition, $g$ is the surface gravity, $\xi$ is the elemental abundance, $\sigma(\lambda)$ is the wavelength dependent absorption cross-section, $P_{o}$ is the pressure at the reference zero altitude, $\tau_{\mathrm{eq}}$ is the optical depth at the transit radius, shown to be approximately constant by Lecavelier des Etangs et al. (2008), and $R_{\mathrm{P}}$ is the radius of the planet. This relationship assumes hydrostatic equilibrium and the ideal gas law.
\citet{lecavelier08} show that the temperature can be inferred from the slope of the line by differentiating Equation (\ref{eqn_guillot2}):

\begin{equation}
\frac{\partial z}{\partial \lambda}=\frac{kT}{\mu g}\frac{\partial (\ln \sigma(\lambda))}{\partial \lambda} \Rightarrow T=\frac{\mu g}{k}\left(\frac{\partial (\ln \sigma(\lambda))}{\partial \lambda}\right)^{-1} \frac{\partial z}{\partial \lambda} .
\label{eqn_guillot_diff}
\end{equation}
In the case of the sodium doublet absorption, measuring the slope of the spectral absorption depth profile in the actual data can be used to derive temperatures for different parts of the profile, and thus temperatures for different altitudes where the absorption takes place: deeper in the atmosphere from the broader parts of the lines and at higher altitudes from the narrower parts of the lines (e.g. \citealt{vidalmadjar11}). Any non-constant slope of the measured spectral absorption depth profile indicates that the profile probes more than one temperature regime.

For HD~189733b, we assume $g=2141$~$\mathrm{cm}\mathrm{s}^{-1}$ and an atmosphere composed of mainly hydrogen and helium (85 per cent H, 15 per cent He), where $\mu=2.3 \times 1.6726 \times 10^{-24}$~g. The sodium abundance, $\xi$, planetary radius, $R_{\mathrm{P}}$, and the reference pressure, $P_{o}$, do not affect the slope of the model spectral absorption profiles and hence the temperatures. We arbitrarily set these values to $\xi_{\mathrm{Na}}/ \xi_{\mathrm{H}}=1.995 \times 10^{-6}$, a solar abundance \citep{lodders03}, $R_{\mathrm{P}}=1.138$~R$_{\mathrm{J}}$, and $P_{o}=150$~mbar. \citet{lecavelier08} show that $\tau_{\mathrm{eq}}=0.56$. To determine $\sigma(\lambda)$ for the sodium doublet, each line was modelled as a Voigt profile, where

\begin{equation}
\sigma(\lambda) = \frac{\sigma_{o}}{\Delta \nu_{D} \sqrt{\pi} }H = \frac{\pi e^2}{m_{e}c}\frac{f}{\Delta \nu_{D} \sqrt{\pi} } H ,
\end{equation}
and $\sigma_{o}$ is the absorption-cross section at the centre of the relevant line, $\Delta \nu_{D}$ is the Doppler width, $H$ is the Voigt function, which includes functions for natural, Doppler and pressure broadening, $e$ is the electronic charge, $m_{e}$ is the mass of the electron, and $f$ is the absorption oscillator strength. We used $f=0.6405$ for the $D_{2}$ line and $f=0.3199$ for the $D_{1}$ line (Steck, 2010)\footnote[2]{See http://steck.us/alkalidata.}. The wavelength-dependent cross-section, $\sigma(\lambda)$, was calculated separately for the D$_{1}$ and D$_{2}$ lines, and the combined profile was worked out using $\sigma(\lambda)=\sigma(\lambda)_{\mathrm{D1}}+\sigma(\lambda)_{\mathrm{D2}}$. The Doppler width is given by  $\Delta \nu_{D} = \nu_{o}/c \sqrt{2kT/\mu_{Na}}$,
where $\nu_{o}$ is the central frequency of the doublet being computed, $c$ is the speed of light and $\mu_{Na}$ is the mean molecular weight of sodium, $23 \times 1.6726 \times 10^{-24}$~g. We used the central wavelengths from Steck (2010), with $\lambda_{o_{D1}}=5895.98 $~\AA $ $ and $\lambda_{o_{D2}}=5890.00 $~\AA. These also match the central wavelengths observed in the data. 

We used Equation (\ref{eqn_guillot2}) to calculate a spectral absorption depth profiles using $\sigma(\lambda)$ and fitted these to each region of the lines with $T$ as a free parameter. To allow us to fit each region locally and account for global changes in abundance, we also fitted the profiles with a free altitude parameter, $z_o$, which allows the profile to shift up and down. 

\begin{figure*}
\centering
\includegraphics[width=8cm]{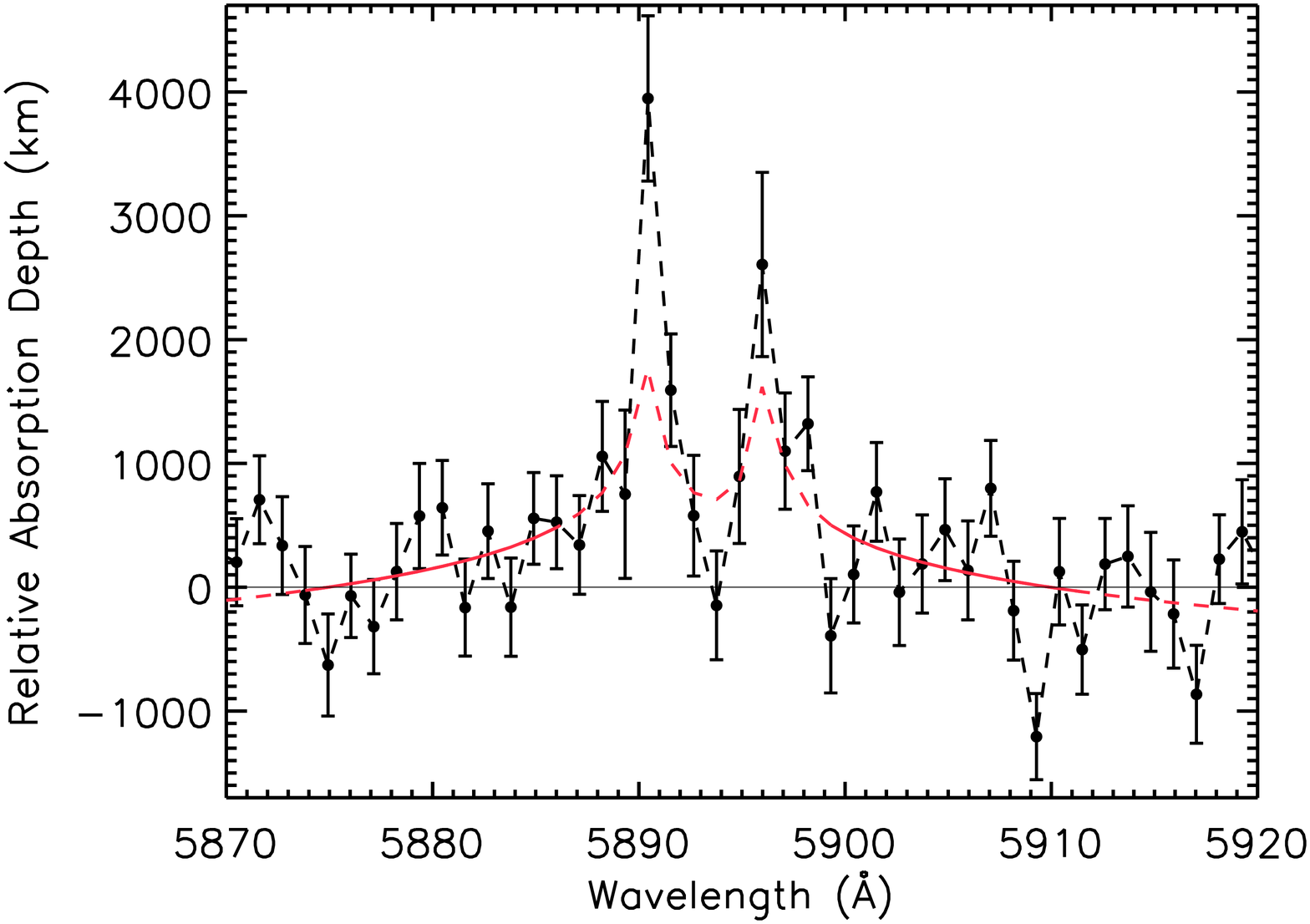}
\includegraphics[width=8.05cm]{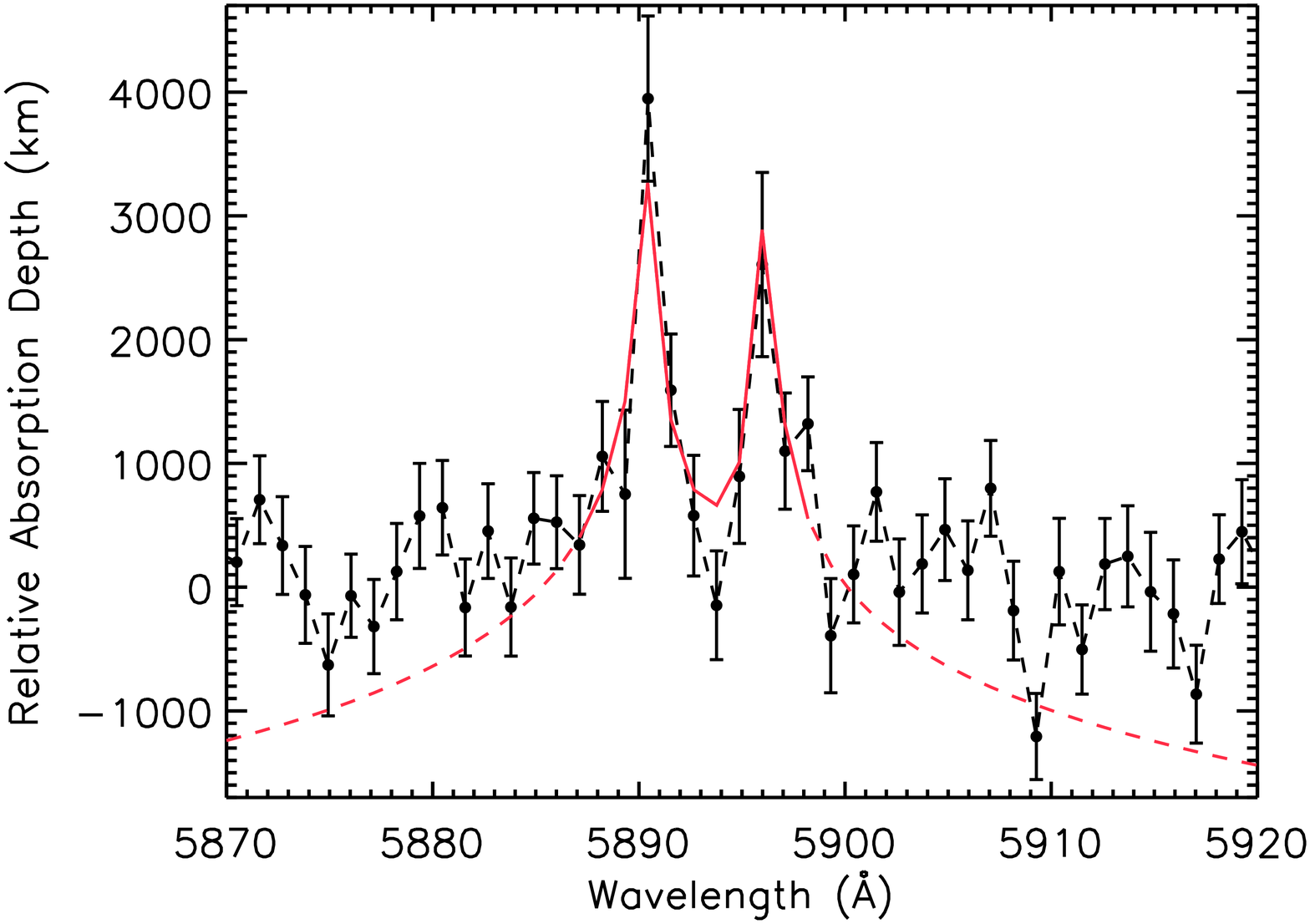}
\caption{Plots showing isothermal model fits to the different wavelength regions of the data. The best fitting models are shown in red, binned to the instrument resolution. \textit{Left: }Fit to `line wings' (5874-5886 \AA $ $ and 5899-5912 \AA), which probes atmospheric regions less than $\sim 500$~km above the reference altitude. The best fitting temperature is $1280 \pm 240$~K, a temperature that is similar to the temperature obtained by fitting the broad-band continuum absorption with Rayleigh scattering at the same wavelength range. \textit{Right: }Fit to the `line cores', at 5887-5898 \AA. This wavelength region probes higher atmospheric regions, greater than $\sim 500$~km above the continuum. The temperature is found to increase to $2800 \pm 500$~K.}
\label{fig_line_temps}
\end{figure*}

We find two temperature regions. Fitting the whole Na profile with one temperature gives a $\chi^2$ of 55.5 and $T=2000$~K for 35 DOF with a BIC of 59.0. However, fitting for two temperature regimes, one at low altitudes probed by the `line wings' and one at high altitudes probed by the `line cores' results in a superior fit, with a BIC of 44.6. The signal in the `line wings' (5874-5886 \AA $ $ and 5899-5012 \AA) of the Na~I~D lines is significant at the 2.9 sigma level added over the whole wavelength region for all visits, indicating a likely detection separate from the `line cores' (5887-5898 \AA). The best fits are shown in Figure~\ref{fig_line_temps}, with the heights in km being relative to the continuum and showing the atmospheric regions probed by each part of the spectral lines.

The best fit to the `line wings' of the Na~I~D feature is $1280 \pm 240$~K, with $\chi^2=28.0$ (24 DOF). This region approximately covers altitudes of less than 500~km above the continuum, which is within the same vertical region in the atmosphere as the blueward rise in absorption depth observed in the near-UV by \citet{sing11}, who found a temperature of $2100 \pm 500$~K. This marginally suggests that another mechanism may be required to explain the excess absorption in the blue region of the spectrum, though the temperatures agree at the 1.5$\sigma$ level. The best fitting temperature for the Na~I `line cores' was found to be $2800 \pm 400$~K, with $\chi^2=11.0$ for 11 DOF. This region corresponds approximately to altitudes of 500-4000~km above the continuum.

We fit the data neglecting pressure broadening. Including pressure broadening in the fit with pressures of 60-150~mbar for the line wings (see Section \ref{haze_mgsio3}) has very little effect, with the best fitting temperature increasing by $\sim 20$~K and the $\chi^2$ increasing to 32.6 for 24 DOF. Pressure broadening does not affect the innermost core of the line.

In order to consider the spectral dispersion in the fits, we moved the centres of our fitted sodium model by one pixel (0.54 \AA) bluewards and redwards of the centre values and found that the `line core' temperatures changed by only $\sim 200$~K. We also calculated the Doppler broadening of the line that should result from the planet's motion about the star, which could lower the temperatures derived from the line shape. We found that, for the phase of the observations, the maximum blueshift (during ingress) and redshift (during egress) is $< 0.7$ \AA, which is below our instrumental resolution.

We further tested the effect of adding Rayleigh scattering at the lowest altitudes by combining the cross-sections for the sodium doublet and a Rayleigh signal. Keeping the `line core' temperature fixed at 2800~K and fitting the `line wings' with a sodium and Rayleigh signal gives a $\chi^2$ of 35.6 instead of 28.0 in the sodium only case, indicating that a Rayleigh component to the signal is likely negligible. Since the temperature derived from the Rayleigh slope by \citet{lecavelier08} is the same as the fit obtained here by assuming that the whole signal in the wings is due to sodium, the presence of a Rayleigh component in our signal does not affect our calculated atmospheric $T$-$P$ profile.

\begin{figure}
\centering
\includegraphics[width=8.7cm]{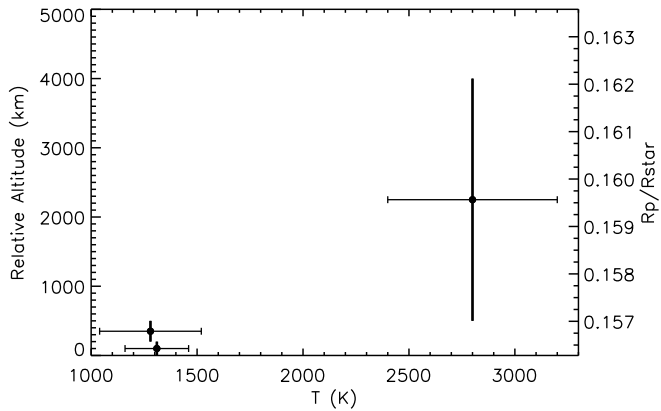}
\caption{Temperature-altitude profile for HD~189733b. The altitude values shown are binned to the \textit{HST} STIS resolution, relative to the continuum level. The thick vertical lines indicate isothermal atmospheric layers. The horizontal error bars show the range of temperatures that match the spectral absorption depth profile for the given atmospheric region. The lowermost point is the temperature derived from the Rayleigh continuum slope by \citet{lecavelier08}. The temperature rise with increasing altitude that we observe occurs much higher than the stratosphere.}
\label{t-z}
\end{figure}

The resulting temperature-altitude profile is shown in Figure~\ref{t-z}, where the altitudes correspond to the approximate atmospheric height ranges probed by each of the fitted wavelength regions of the doublet. This profile includes the previously measured temperature of $1340 \pm 150$~K found by \citet{lecavelier08} for the Rayleigh continuum slope in the sodium region, at altitudes of $z=0$-$200$~km. 

\section{Discussion}
\label{discussion}

\subsection{Interpretation of the Narrow Absorption Lines}

\begin{figure}
\centering
\includegraphics[width=8.5cm]{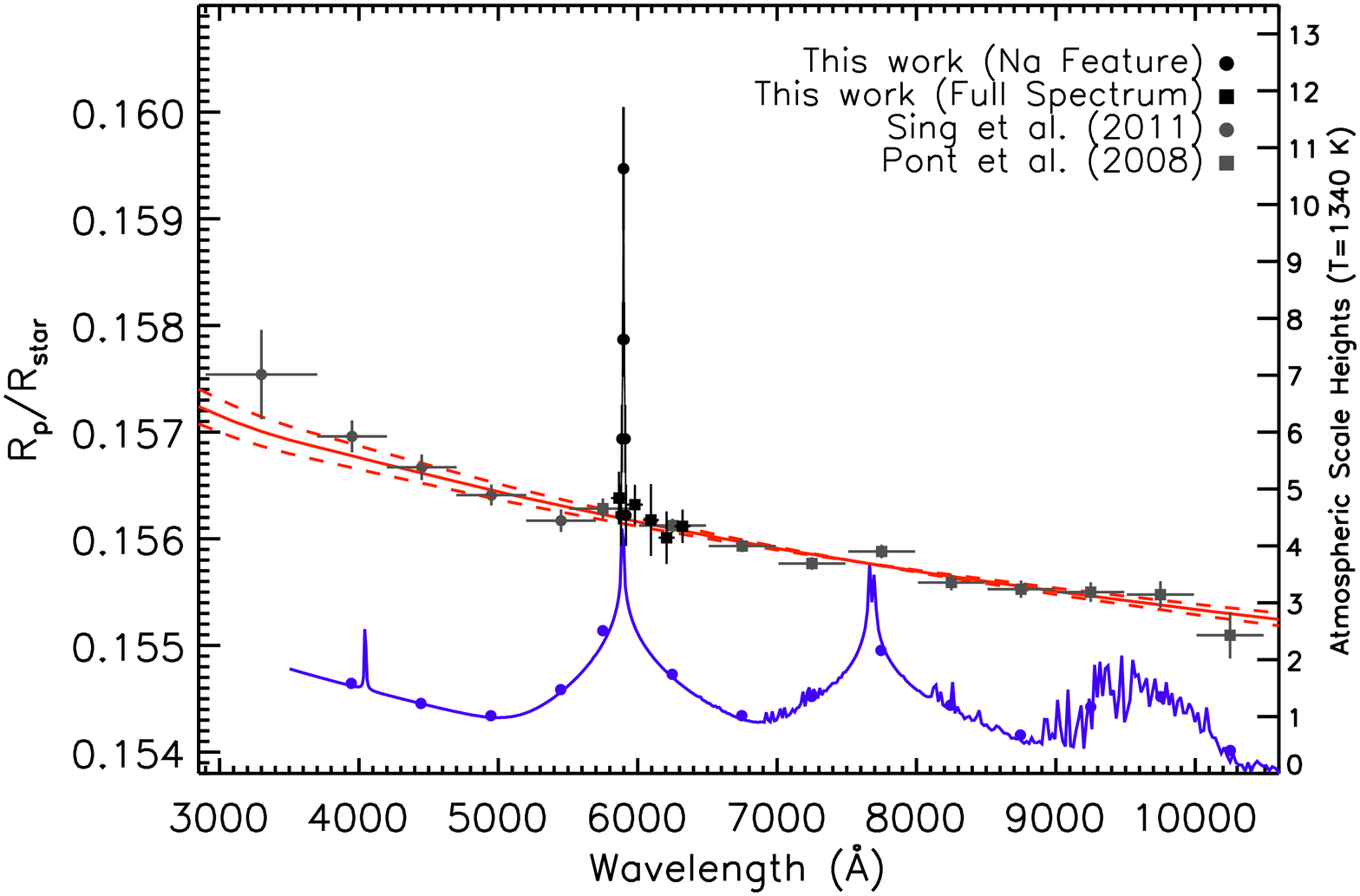}
\caption{Spectrum over the G750M band (black squares) compared with previous work. Also shown are points from the G750M spectrum at 2, 5, 12 and 18 \AA $ $ from the centre of the doublet (black circles). The red lines show the Rayleigh scattering prediction from the ACS data \citep{lecavelier08}. The blue line shows a model assuming a haze-free atmosphere from \citet{fortney10} at solar Na~I abundance, normalised to the radius at infrared wavelengths from \citet{agol10}. Blue circles show model  values binned to the G430L and ACS resolution. A colour version is available in the online journal.}
\label{fig_haze_sodium}
\end{figure}

Figure~\ref{fig_haze_sodium} shows the previously observed broad-band spectrum \citep{pont08, sing11} compared to a model spectrum assuming a haze-free atmosphere \citep{fortney10} and a solar Na~I abundance. The observed spectrum lacks the broad alkali line wings and H$_{2}$O signatures. The broadband data are binned to $\sim 500$~\AA, showing that the equivalent widths of those bins are not significantly higher than the featureless spectrum. Figure~\ref{fig_haze_sodium} also shows our measured spectrum for the G750M band as well as points from the medium resolution spectrum at 2, 5, 12, and 18 \AA $ $ from the centre of the doublet, confirming a narrow Na~I feature. The previous broad-band spectroscopy would not have had sufficient spectral resolution to resolve such a narrow feature unambiguously. The 3 observations from STIS G750M, STIS G430L \citep{sing11} and ACS HRC \citep{pont08} all independently agree that there are no observable broad Na~I line wings.

The species responsible for the Rayleigh scattering signature observed in HD~189733b is unknown, but the signature should place constraints on the composition of the scattering species \citep{lecavelier08,sing11}. Rayleigh scattering dominates only for particle sizes much smaller than the wavelength and where the scattering efficiency is dominant over the extinction efficiency. H$_{2}$, which is abundant in the atmosphere, is much smaller than the observed wavelengths. A narrow Na~D doublet profile can be explained by a low sodium abundance which can hide the Na~I line wings beneath the H$_{2}$ Rayleigh signature. 

Alternatively, if the abundance is solar or above, a high-altitude material could cover the lower regions of the line, allowing only the upper regions to be observed, where pressures are low and there is no observable pressure broadening of the feature. If the Rayleigh scattering is from a haze of condensate particles, then the condensate must be transparent enough for the absorption to be negligible relative to scattering, which rules out many condensates such as MgSiO$_{4}$ and oxygen-deficient silicates \citep{lecavelier08}. One possible species is MgSiO$_{3}$. Assuming a uniform haze layer with a single size component, the observed broad-band data from STIS and NICMOS constrain the particle size to be in the range 0.01-0.03~$\umu$m by requiring that the molecule scatters at all observed wavelengths. 

The narrow range of particle sizes needed to be consistent with the observations shows that it is potentially difficult to explain the broad-band Rayleigh signature over the 4000-18700~\AA $ $ range with scattering by a haze of condensates of constant grain size. It is more likely that a condensate would vary in grain size with depth, and that there could be more than one absorber (e.g. aerosols) but since the scattering species is so unknown, we present only a representative $T$-$P$ profile based on these grain sizes for a single absorber.

\subsection{Temperature-Pressure Profile}
\label{temperature_pressure}

The temperatures of the upper atmosphere probed by the optical observations increase with increasing altitude. For our scale heights, we assume that most of the hydrogen is in the form of H$_{2}$. At higher altitudes, \citet{yelle04} suggest that the scale height could increase further due to the change in dominance of hydrogen from H$_{2}$ to H, which would decrease the mean molecular weight, $\mu$. A lower mean molecular weight would extend the line without having to increase the temperature. \citet{garciamunoz07} show that H dominates over H$_{2}$ only for pressures lower than $P=10^{-8}$ bar. \citet{moses11} also estimates that photodissociation of H$_2$ occurs sharply at $10^{-8}$~bar, although this pressure depends on where in the atmosphere the stellar radiation is deposited and could be at higher pressures. Our Na~I line cores (highest altitudes) occur at $\sim r/R_{\mathrm{P}} \sim 1.06$, where H$_{2}$ should dominate.

Our measurements probe atmospheric regions well above the stratosphere, which could indicate that we are observing the base of the atmospheric thermosphere. To compare our temperature profile with theoretical models, we generate a $T$-$P$ profile, using our derived temperatures to work out the corresponding atmospheric scale heights, and hence the pressure structure of the atmosphere relative to the reference level. We assume a well-mixed atmosphere where the scale heights derived from the sodium line will be representative of the atmosphere. 

The measured spectral absorption depths are only relative to the absorption depth in the reference bands defined in Section~\ref{na_spectral_profile} and hence the derived $T$-$z$ profile can only be converted into a $T$-$P$ profile if the pressure at the reference altitude is known (see discussion in \citealt{lecavelier08b}). Determining a reference pressure requires the species responsible for the continuum absorption in the reference bands and its abundance to be known. In the case of HD~209458b, the identification of Rayleigh scattering by H$_{2}$ provided the required abundance information \citep{vidalmadjar11}. Since the continuum absorption in HD~189733b is due to an unknown species of unknown abundance, the inferred pressure-altitude relationship is not well constrained and can shift up or down depending on the reference pressure. 

We used a pressure profile based on a reference pressure of $150$~mbar at $z=0$ ($R_{\mathrm{P}}=0.15628$) to determine a pressure-altitude relationship assuming H$_{2}$ as the scattering species. Any other Rayleigh scattering species will be at higher altitudes, meaning that the pressure profile based on H$_{2}$ as the scattering species is an upper limit. The $T$-$P$ profile was arbitrarily shifted to lower pressures to consider $T$-$P$ profiles that result from different possible scattering species. \citet{lecavelier08} show that a haze of MgSiO$_{3}$ grains of sizes of 0.01-0.03~$\umu$m at a temperature of 1340~K will be at pressures between $10^{-3} < P < 10^{-5}$~bar for solar magnesium composition. 

\subsubsection{The Terminator T-P Profile of HD~189733b}   
\label{haze_mgsio3}

\begin{figure*}
\centering
\includegraphics[width=8.7cm]{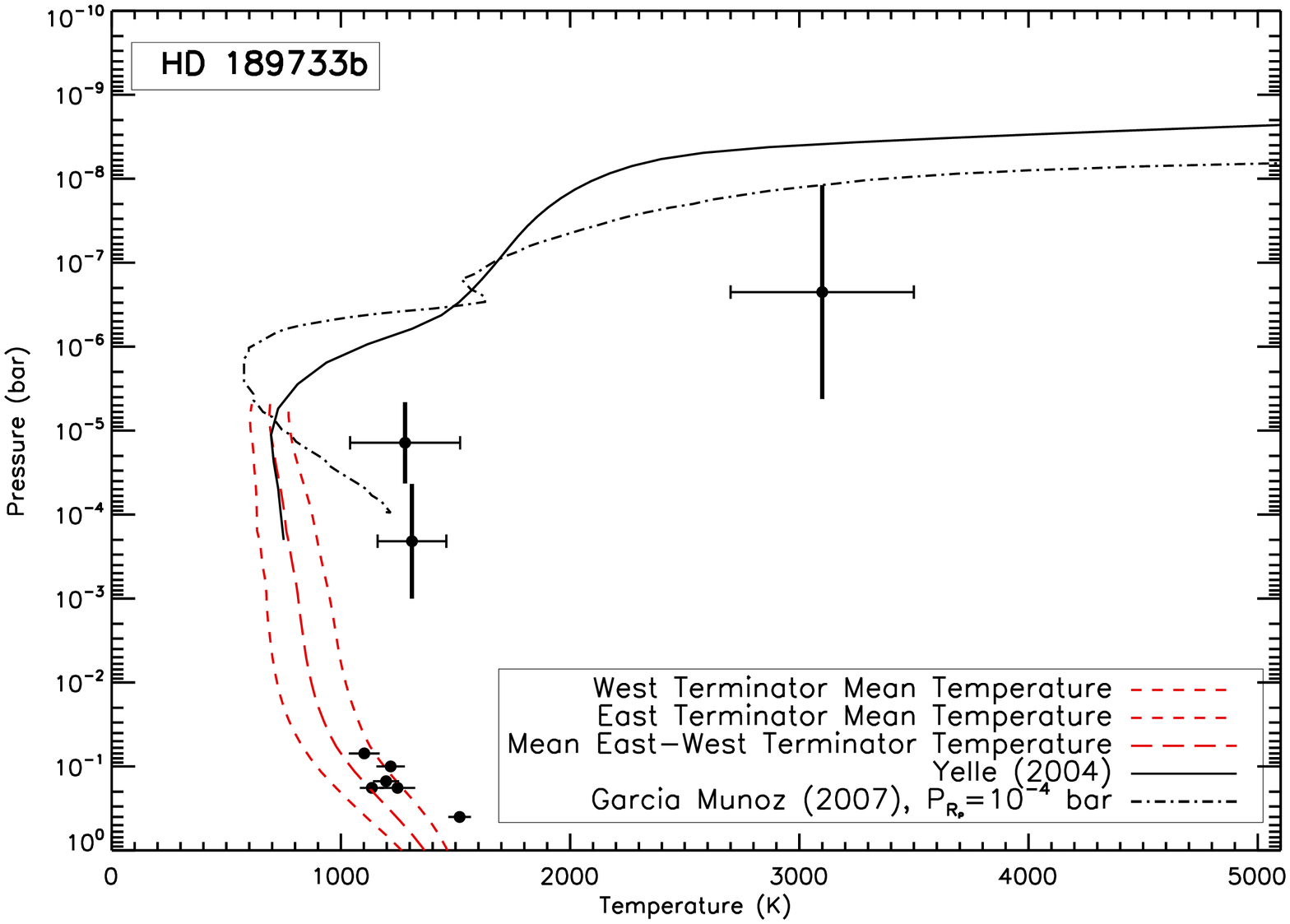}
\includegraphics[width=8.7cm]{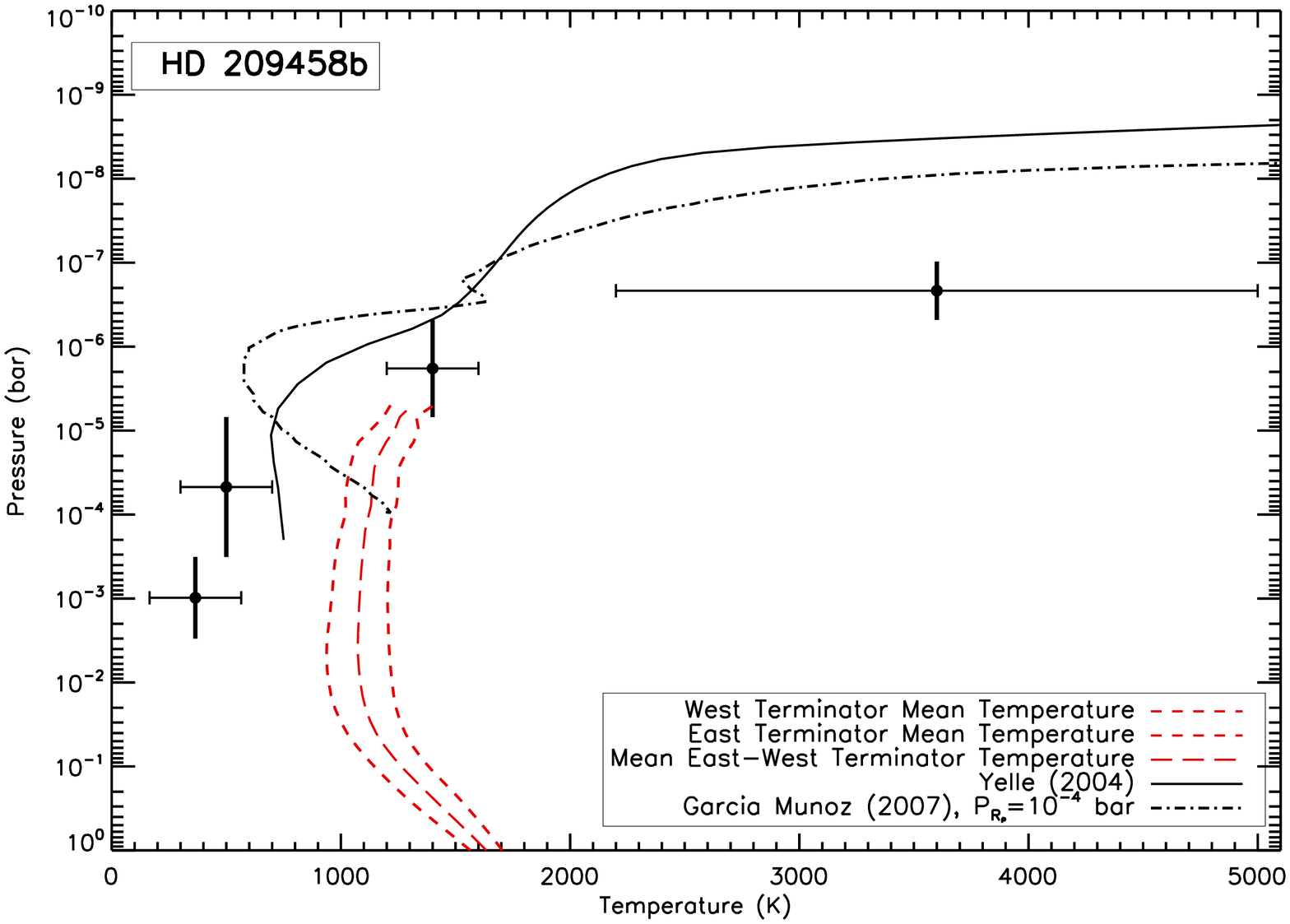}   
\caption{\textit{Left: }$T$-$P$ profile for HD~189733b based on the assumption that the Rayleigh scattering signature is due to high-altitude atmospheric haze at a reference pressure of $10^{-4}$~bar. The temperatures are plotted in black for each pressure range (thick vertical bars). The thick vertical bars do not include the uncertainty in reference pressure. The data include the 1340~K temperature and pressure derived for the continuum by \citet{lecavelier08} combined with the 1280~K temperature measured for the near infrared continuum by  and \citet{sing09}. The plot also shows lower altitude points from \citet{knutson07b,knutson09}, \citet{charbonneau08} and \citet{deming06}. The temperatures measured by \citet{charbonneau08} and \citet{deming06} were adjusted to the terminator temperature by \citet{heng11}. $T$-$P$ model profiles are shown for the lower altitude regions with red dashed lines \citep{fortney10} and for the higher altitudes we show models for the upper atmosphere of HD~209458b for comparison, where the black solid line is the $T$-$P$ profile from \citet{yelle04} and the dot-dashed line is from \citet{garciamunoz07}. \textit{Right: } $T$-$P$ profile for HD~209458b using the measurements from \citet{vidalmadjar11b}, plotted as black points. Shown with red dashed lines are models for the lower atmosphere \citep{showman09}. The solid black line is a model from \citet{yelle04} and the dot-dashed line is from \citet{garciamunoz07}.}
\label{tp_mgsio3}
\end{figure*}

Figure~\ref{tp_mgsio3} shows the $T$-$P$ profile derived assuming that the Rayleigh signature is due to a high-altitude silicate haze. The reference pressure is $10^{-4}$~bar, representative of the $10^{-3}-10^{-5}$~bar range that fits the broad-band observations. Depending on the species, this profile could shift significantly. In the case assuming instead that H$_{2}$ is the scattering species, the derived pressures would be 150~mbar for the reference level, 65 to 6.6~mbar for the lower region of the observable line at $1280 \pm 240$~K and 6.5 to 0.02~mbar for the upper region of the line at $2800 \pm 400$~K. We also show the temperatures and pressures observed for lower altitudes (higher pressures) from \citet{knutson07b,knutson09}, \citet{charbonneau08} and \citet{deming06}. The temperatures in the last two works were adjusted to the terminator temperature by \citet{heng11} and these are the values we use. We also combine the temperature measurement from \citet{sing09} with that from \citet{lecavelier08} since they are both derived from fits to the broad-band Rayleigh slope. The models shown for comparison to HD~189733b are from \citet{fortney10}, \citet{yelle04} and \citet{garciamunoz07}. Both upper atmospheric models were computed for HD~209458b. The vertical bars which indicate pressure range in the figure for HD~189733b do not include the uncertainty in the pressure reference level. They indicate the altitude range over which we match each temperature.

The pressures where we observe the strong temperature rise are at least an order of magnitude higher than the pressures of the base of the thermosphere found in the models of the upper atmosphere of HD~209458b, assuming a high-altitude haze as a reference level. This difference increases even further, by 3 orders of magnitude, if we assume H$_{2}$ as the scattering species. The discrepancy between our data and the models could indicate that we are observing a lower-altitude temperature rise rather than the temperature rise characteristic of the base of the thermosphere, although the mechanism for such a deeper temperature rise is unknown. Alternatively, the discrepancy could be because the scattering species is higher in the atmosphere than our MgSiO$_{3}$ example case. Our pressures can shift by an order of magnitude assuming an MgSiO$_{3}$ haze, and could shift even more if another species is responsible for the scattering continuum. 

 It is important to note that, if our pressure scale shifts significantly, our assumption that H$_2$ is the dominant atmospheric species may become invalid and, in this case, the 2800~K temperature would be an overestimate. Conversely, if the abundance decreases significantly over the ranges where we measure the individual temperatures, for example due to ionisation, this will cause us to underestimate the temperature. Further theoretical work is required to better understand the overall properties of the atmosphere, especially as the models for the upper atmosphere of HD~209458b may not adequately describe HD~189733b.

\subsubsection{Comparing the T-P Profiles of HD~209458b and HD~189733b}

Figure~\ref{tp_mgsio3} also shows the $T$-$P$ profile derived for HD~209458b by \citet{vidalmadjar11b}, which was calculated using the same method as in this paper, of fitting temperatures to the spectral absorption depth profile. It is an update to the previous profile for HD~209458b by \citet{vidalmadjar11}, which was determined by fitting temperatures to the integrated absorption depth profile, where measurements of the slope of the line can be compromised by dilution effects of increasing bandwidth around an unresolved line. The authors find that the two analyses are equivalent for regions where the line is resolved and firmly detected, and their conclusions remain unchanged, but the error analysis is non trivial due to all of the points in such a profile being co-dependent. For comparison to the HD~209458b data, the models of \citet{showman09}, \citet{yelle04} and \citet{garciamunoz07} are shown. Comparing the calculated $T$-$P$ profiles for both planets to the models shows that the model profiles predict pressures lower than those derived from the observations even for HD~209458b at the thermobase. There are also no equivalent models for the upper atmosphere for HD~189733b, since existing models use solar-type stellar spectra. The models of \citet{moses11} make allowances for the reduced emission at wavelengths longer than 2830~\AA, but the thermosphere is placed at a specific pressure, based on the models of \citet{garciamunoz07}. Since there is no work at present that is able to model all of the factors at the base of the thermosphere or at low pressures for HD~189733b, we cannot draw any meaningful comparisons with existing models. However, it is hoped that these observations will be able to constrain future theoretical work. 

The most useful comparison to draw currently is between the two $T$-$P$ profiles calculated from observations. To compare the two planets, we assume that our observed temperature rise indicates a detection of the base of the thermosphere, although we cannot be sure that the temperature continues to rise above our observed altitude regions. Both planets have measured escaping atmospheres \citep{vidalmadjar03,lecavelier10}, which should be associated with extremely high temperatures at high altitudes. 

The shape of the derived $T$-$P$ profile for the upper most regions of HD~189733b looks similar to the HD~209458b profile. However, this conclusion has to be tentative as the infrared data points are not tied to our pressure scale, so the overall $T$-$P$ profile over all observed pressures would look different if the upper atmospheric region of the profile were shifted. It would be very interesting to resolve the spectral absorption depth profile further and determine more accurately the shape of the $T$-$P$ profile. The speed with which the upper atmosphere heats up with increasing altitude could give clues about any absorbing or reflecting material high in the atmosphere.


\subsection{The Relative Atomic Sodium Abundance}
\label{water_abundance}

Abundance measurements of identified spectral features can be determined from transmission spectra by rearranging Equation~(\ref{eqn_guillot2}). The relative abundance of two species is given by (also shown by \citealt{desert09}):

\begin{equation}
\frac{\xi_{1}}{\xi_{2}}=\frac{\sigma_{2}}{\sigma_{1}}e^{\{(z_{1}-z_{2})/H\}} ,
\label{eqn_abundances}
\end{equation}
where $\xi_{1}$ and $\xi_{2}$ are the abundances, $\sigma_{1}$ and $\sigma_{2}$ are the wavelength dependent cross-sections at $\lambda_{1}$ and $\lambda_{2}$, and $z_{1}$ and $z_{2}$ are the altitudes of the observed spectral features at $\lambda_{1}$ and $\lambda_{2}$ of species 1 and 2 respectively. All other constants cancel out assuming that both spectral features occur in the same temperature regime. 

We compared our sodium absorption depth with the radius observed at 8~$\umu$m by \citet{agol10}, assuming that this feature is due to absorption from atmospheric water. The results from secondary eclipse observations have detected a water absorption signal at this wavelength \citep{grillmair08}, leading us to believe that water is likely to be the source of the 8~$\umu$m feature we see in transmission as already envisaged by \citet{desert09}. We used Equation (\ref{eqn_abundances}) along with $\sigma_{\mathrm{H_{2}O}}=2 \times 10^{-20}$~cm$^2$ at a wavelength of 8~$\umu$m \citep{desert09} and a radius for the water feature at 8~$\umu$m of $R_{\mathrm{P}}/R_{\star}=0.15531$ \citep{agol10}. This was compared to the radius in the region 10-18~\AA $ $ from the centre of the Na~I doublet, which was $R_{\mathrm{P}}/R_{\star}=0.1572 \pm 0.0002$. This wavelength region has a similar temperature to the continuum and has absorption cross-section $\sigma_{\mathrm{Na}}=3.5 \times 10^{-20}$~cm$^2$. Since this is not a differential measurement, we used only the fitted radius from visit 7, which is has the smallest contamination from occulted stellar spots of all the visits, rather than an average of all 3 visits. The un-occulted spot correction used for the G750M band was 1 per cent, and the un-occulted spot correction used for the 8~$\umu$m band was 0.2 per cent (see \citealt{sing11}).

The solar abundance ratio of sodium to water is $\xi_{\mathrm{Na}}/\xi_{\mathrm{H_{2}O}}[\mathrm{solar}] =6.63-9.95 \times 10^{-3}$ or $\ln(\xi_{\mathrm{Na}}/\xi_{\mathrm{H_{2}O}}[\mathrm{solar}])=-5.0$ to $-4.6$ \citep{lodders03,loddersfegley02,sharpburrows07}. For HD~189733b, we calculated values of $\xi_{\mathrm{Na}}/\xi_{\mathrm{H_{2}O}}[\mathrm{189}]$ which were $\sim 100$ times higher ($\ln(\xi_{\mathrm{Na}}/\xi_{\mathrm{H_{2}O}}[\mathrm{189}])=+1.3 \pm 0.75$), indicating a super-solar abundance ratio.

\section{Summary and Conclusions}

We have detected sodium in the atmosphere of HD~189733b with $9 \sigma$ confidence, confirming the previous ground-based measurement, and have improved the precision of the measured absorption level by almost a factor of 3. Our improved resolution has enabled us to measure the sodium doublet absorption profile. We confirm the presence of a very narrow Na~I doublet feature, which can be explained by a high-altitude haze obscuring broad Na, K and H$_{2}$O features, or a low Na~I abundance hiding the wings beneath an H$_{2}$ scattering signature.

We use the Na~I spectral absorption depth profile to determine the temperature as a function of altitude and find that two temperature regimes are required to explain the observed profile. The temperature rises with increasing altitude over the regions we have measured, indicating a likely detection of the planet's thermosphere. A thermosphere is expected from models and has been detected in HD~209458b. So far, only two planets have measured upper-atmosphere temperature profiles, but it is reasonable to assume that most hot Jupiters will exhibit hot thermospheres, due to their closeness to their host stars and the amount of radiation they receive in their upper atmospheres. 

We calculated a resulting $T$-$P$ profile assuming a high-altitude haze as the reference level, based on MgSiO$_{3}$ as the scattering species. It is possible that condensates such as MgSiO$_{3}$ could sublime at temperatures of $\sim 1300$~K or even lower temperatures at the low pressures sensed with the visible data (e.g. \citealt{moses11}), which could indicate that another species is responsible for the observed signature. The pressures are 3 orders of magnitude higher assuming H$_{2}$ Rayleigh scattering as the reference level, indicating how uncertain the pressure scale is. Additionally, the pressure scale with altitude will change if the sodium abundance is not solar.

We find that we cannot constrain the Na~I abundance without knowing the species responsible for the continuum signature and hence our reference level, but comparison with a feature at 8~$\umu$m indicates that the relative abundance of sodium compared to water is much greater than solar.

We have measured only the second upper-atmospheric temperature-altitude profile for an extrasolar planet. The method described in this paper could be applied to other planets, leading to further understanding of their upper atmospheres and how they interact with their host stars. The line profile gives information about whether high-altitude optical absorbers are present, allows us to determine an atmospheric temperature structure and will be invaluable for studying the atmospheric escape mechanism. It can also allow us to place some constraints on the elemental abundances if a reference pressure can be determined. It is thought that there are at least two classes of hot Jupiters, but more measurements are required to determine whether HD~189733b and HD~209458b are representative of the different classes. The case of WASP-17b, which has an absorption profile that is yet different again \citep{wood11}, suggests a range of properties within these classes, with an even narrower line profile suggesting an optical haze appearing higher in the atmosphere than in HD~189733b or an atomic sodium abundance lower than in HD~189733b. With regard to HD~189733b specifically, an optical secondary eclipse measurement would show us whether there is reflection or scattering from absorbers high in the atmosphere, which will help constrain the nature of such absorbers. Additionally, high resolution ground based spectra are required to resolve this inner-most line core region, and extend the T-P profile to lower pressures and higher altitudes.

\section*{Acknowledgments}

We thank David Charbonneau and Ron Gilliland for their last minute idea of the real time \textit{HST} pointing correction which saved the data quality of visit 3. We also thank Jonathan Fortney for providing his model atmospheres and Adam Showman for providing his model $T$-$P$ profiles. We thank our reviewer, Ignas Snellen, for pointing out problems with using the integrated absorption depth profile to determine atmospheric temperature. This lead us to the idea of using the spectral absorption depth profile, which has greatly improved the paper. This work is based on observations with the NASA/ESA Hubble Space Telescope. This research has made use of NASA's Astrophysics Data System, and components of the IDL astronomy library. C.M. Huitson acknowledges support from STFC. G.E. Ballester acknowledges support by NASA through grant HST-GO-11576-01-A to University of Arizona from STScI. 

\bibliography{Na_189.bib} 

\label{lastpage}

\end{document}